\documentclass[12pt,titlepage]{article}
\newcommand{\pzzm}{\left( \begin{array}{cc} 1\!\!1 & 0 \cr 0 & -1\!\!1 \end{array} \right)}
\newcommand{\pzzpm}{\left( \begin{array}{cc} 1\!\!1 & 0 \cr 0 & \pm1\!\!1 \end{array} \right)}
 
\newcommand{\zpmz}{\left( \begin{array}{cc} 0 & 1\!\!1 \cr -1\!\!1 & 0 \end{array} \right)}
\newcommand{\zppmz}{\left( \begin{array}{cc} 0 & 1\!\!1 \cr \pm 1\!\!1 & 0 \end{array} \right)}

\newcommand{\isy}{\left( \begin{array}{ccccc} \! 0 \!  & \! 1 \! & & & \cr \! -1\!  & \!  0\!  & & & \cr & &\! \ddots \! & & \cr & & & \! 0\!  & \! 1 \! \cr & & & \! -1\!  & \! 0 \! \end{array} \right)}
\newcommand{\zppz}{\left( \begin{array}{cc} 0 & 1\!\!1  \cr 1\!\!1 & 0 \end{array}\right)}

\newcommand{\ISY}{\left( \begin{array}{cc} \!\!\! \begin{array}{cc} 0 & 1\!\!1 \cr -1\!\!1 & 0 \end{array} \!\!\! & \!\!\! \!\!\! \cr
                                            \!\!\! \!\!\!  &  \!\!\! \begin{array}{cc} 0 & 1\!\!1 \cr -1\!\!1 & 0 \end{array} \!\!\! \end{array}\right) }
\newcommand{\ISy}{\left( \begin{array}{cc} \!\!\! \begin{array}{cc} 0 & 1\!\!1 \cr -1\!\!1 & 0 \end{array} \!\!\! & \!\!\! \!\!\! \cr
                                            \!\!\! \!\!\!  &  \!\!\! \begin{array}{cc} 0 & -1\!\!1 \cr 1\!\!1 & 0 \end{array} \!\!\! \end{array}\right) }
\newcommand{\XISY}{\left( \begin{array}{cc} \!\!\! \!\!\! & \!\!\! \begin{array}{cc} 0 & 1\!\!1 \cr -1\!\!1 & 0 \end{array} \!\!\!  \cr
                                                  \!\!\! \begin{array}{cc} 0 & 1\!\!1 \cr -1\!\!1 & 0 \end{array} \!\!\! & \!\!\! \!\!\! \end{array}\right) }

\newcommand{\SX}{\left( \begin{array}{cc} \!\!\! \begin{array}{cc} 0 & 1\!\!1  \cr 1\!\!1 & 0 \end{array} \!\!\! & \!\!\! \!\!\!  \cr \!\!\! \!\!\! & \!\!\!  \begin{array}{cc} 0 & 1\!\!1  \cr 1\!\!1 & 0 \end{array} \!\!\!  \end{array} \right)}
\newcommand{\SZ}{\left( \begin{array}{cc} \!\!\!  \begin{array}{cc} 1\!\!1 & 0 \cr 0 & -1\!\!1 \end{array} \!\!\!  & \!\!\! \!\!\! \cr \!\!\! \!\!\!  & \!\!\!  \begin{array}{cc} 1\!\!1 & 0 \cr 0 & -1\!\!1 \end{array} \!\!\!  \end{array} \right)}

\renewcommand{\to}{\rightarrow}
\newcommand{\beq}{\begin{equation}}
\newcommand{\eeq}{\end{equation}}
\newcommand{\bea}{\begin{eqnarray}}
\newcommand{\eea}{\end{eqnarray}}  

\newcommand{\barc}{\begin{array}{c}}
\newcommand{\ear}{\end{array}}
\def\ln{{\rm ln}}

\begin{document}
\thispagestyle{empty}
\begin{titlepage}
\addtolength{\baselineskip}{.7mm}
\thispagestyle{empty}
%
\begin{center}
{\large 
{\bf Random matrices beyond the Cartan classification
}}\\  [11mm]
{
\bf 
Ulrika~Magnea 
} \\  
\vspace{3mm}
{\it Dipartimento di Scienze e Tecnologie Avanzate \\ 
Universit\`a di Piemonte Orientale \\   }
\vskip2mm 
and \\ \vskip2mm
{\it INFN, Sezione di Alessandria \\ \vskip2mm
Via G. Bellini 25/G, I--15100 Alessandria, Italy} \\

{\ }\\ 
{\tt blom@to.infn.it} 
\\ [4mm]
\vspace{15mm}
\end{center}

\centerline{\bf Abstract}

It is known that hermitean random matrix ensembles can be identified with symmetric coset spaces
of Lie groups, or else with tangent spaces of the same. This results in a classification of random matrix ensembles
as well as applications in practical calculations of physical observables. 
In this paper we show that a large number of non--hermitean random matrix ensembles defined
by physically motivated symmetries -- chiral symmetry, time reversal invariance, space rotation
invariance, particle--hole symmetry,
or different reality conditions -- can likewise be identified with symmetric spaces. We give explicit  
representations of the random matrix ensembles identified with lateral algebra subspaces,
and of the corresponding symmetric subalgebras spanning the group of invariance.
Among the ensembles listed we identify as special cases
all the hermitean ensembles identified with Cartan classes of symmetric spaces and the three Ginibre ensembles with complex eigenvalues.

\end{titlepage}

\newpage
\setcounter{footnote}{0}
\section{Introduction}
\label{sec-intro}

Ensembles of random matrices appear with increasing frequency in various branches of
physics. 
In a typical application, an appropriately chosen ensemble of random matrices (hermitean or not) replaces an 
ensemble of physical quantum operators with known global symmetries in problems
involving highly complex, interacting systems. The method, pioneered by Wigner \cite{Wigner} and Dyson \cite{Dyson} over 50 years ago
in a statistical approach to spectra of nuclear energy levels, can produce useful information of a
universal character on the spectrum of eigenvalues of such operators, ranging from Hamiltonians to transfer matrices,
scattering matrices, and Dirac operators. The most characteristic feature of spectra described by 
random matrix theory is the presence of strong geometrical correlations between eigenvalues.
Dyson \cite{DysonSS}, followed by H\"uffmann \cite{Huff}, were the first to 
point out the connection between ensembles of random matrices, subject to certain symmetry conditions, and geometrical manifolds referred to as symmetric spaces. Further work on this was done by Altland and Zirnbauer and by Caselle in references \cite{AZ,MCa,AZC}. In reference \cite{Zsuper} 
the classification of symmetric superspaces in random matrix theory is discussed.

The relative simplicity with which universal spectral characteristics in widely diverse physical
systems can be described in this approach, has led to a huge increase in activity in the past two decades or so,
and the literature in the field by now comprises thousands and thousands of pages (for reviews see 
\cite{reviews}). Initially only hermitean
operators were studied, and ensembles were classified according
to Cartan's scheme of symmetric spaces, using a  
group theoretical viewpoint (for a review of this approach see \cite{SS}). Because of the identification of the set
of random matrix eigenvalues with radial coordinates on a symmetric space, the distinguishing characteristics of the 
matrix ensembles follow naturally via the Jacobian from the underlying geometry of the coset space and the root system
of the associated Lie algebra subspace. This geometrical connection has so far been applied 
mainly in condensed matter physics \cite{MCa, MC,nano}, but the possibility of applications in other branches of physics
should not be overlooked. 

We observe that most of the time, the diffusion operator on the symmetric space,
corresponding to the first Casimir invariant associated to the enveloping algebra, played an important role in these applications. This was the case also in \cite{nano}, where physical measurable quantities of a carbon nanotube were expressed {\it entirely} in terms of quantities characterizing 
the root lattice of the symmetric space on which the transfer matrix lives (for details see section 
\ref{sec-ss}).
This shows that the connection with an
underlying symmetric space is not just a theoretical concept, but can be used for determining {\it physical
observables}.

Lately however, an increasing number of authors employ {\it non--hermitean} random matrix ensembles in a range
of different contexts (for a review with emphasis on applications in quantum chaotic scattering,
see \cite{fyosom1}). 
Except for being interesting in their own right, these new ensembles are relevant to 
unexplored phenomena in 
field theory and condensed matter physics, and appear also in other contexts, for example in the dynamics of neural networks. 
Whereas the range of eigenvalues of a hermitean ensemble is found on the real axis, the 
corresponding support of eigenvalues of a non--hermitean matrix ensemble occupies a two--dimensional 
domain in the complex plane. 
In this paper we take a first step towards clarifying the issue of whether there is 
a corresponding geometrical structure that characterizes the relatively less known non--hermitean "relatives" of 
the known, classified hermitean ensembles, taking as a starting point the paper by Bernard and LeClair \cite{BLC} in which non--hermitean random matrix ensembles were classified according to symmetry.
Except for two new ensembles apparently forgotten in \cite{BLC}, we verify the classification of this reference. The main result of this work is to show that all the resulting ensembles correspond to geometrical symmetric spaces, the classification of which goes beyond the standard Cartan classification of 
{\it hermitean} random matrices 
(we do expect, however, that the non--hermitean ensembles will be identified
with Cartan classes possessing a complex structure). 
As we explain later, we do not claim that the list of symmetric spaces
is complete, nor that it comprises all possible random matrix ensembles. However, it does appear to include many of the purely random matrix ensembles studied in the literature. The most important exception
arises if one adds a fixed matrix to a random ensemble, or imposes an extra condition like the condition
of normality (a matrix $N$ is called normal if $[N,N^\dagger]=0$) on the ensemble. The latter represents
a special case (subset) of certain symmetry classes in our classification.

\subsection{A few applications of non--hermitean random matrix theory}
\label{sec-appl}

As a further motivation to this work, let us give a brief list of physical situations where non--hermitean random matrix ensembles
play a role.

An interesting example occurs in condensed matter physics and is due to Hatano and Nelson
\cite{HatNel}. 
A new phenomenon related to a non--hermitean random Hamiltonian is represented by 
the existence of extended eigenstates of the latter -- referred to as delocalization -- in one and two spatial dimensions.
The delocalization transition can be mapped to the depinning of flux lines from columnar defects in a superconductor
at a critical value of the magnetic field perpendicular to the defects. 

A problem with similar aspects was considered by Guruswamy, LeClair and Ludwig in \cite{Guru},
where disordered fermions in 2d were studied using complex random potentials.

In mesoscopic physics, charge transport in quasi--1d quantum wires is described using the Landauer formalism. A central
object in this approach is the transfer matrix. The 
transmission eigenvalues related to the latter determine the main observable, the conductivity
of the wire. The distribution of transmission eigenvalues is determined, within the random matrix approach, by the solution to a
Fokker--Planck type diffusion equation named after its authors the DMPK (Dorokhov--Mello--Pereyra--Kumar) equation.
In the presence of absorption or amplification in the wire, this equation becomes untractable. However, one can instead study
the reflection properties of the wire in a similar diffusion equation. In this case the scattering matrix of the system is not unitary and it is parametrized
in terms of a non--hermitean matrix \cite{BruceChalk}. 
The same formalism is applicable to multiple scattering of classical waves in a random medium, where one is interested in the statistical properties of the transmitted and reflected waves, and the phenomenon of localization likewise takes place \cite{Be-TitB}. 

Random matrix ensembles with complex eigenvalues are also relevant in the field of quantum chaotic scattering,
where universal aspects of bound state statistics and resonance poles in open systems are described using random matrix
techniques. 
The pioneering work by Verbaarschot, Weidenm\"uller and Zirnbauer
can be found in reference \cite{pio}; see also the works by Fyodorov and Sommers, \cite{fyosom1, fyosom}. There are a good number of published studies in this field, see for example \cite{fyokor}.

One of the possible applications of random
matrix models relevant to QCD involves substituting an appropriate random matrix 
for the Dirac operator (fermion determinant) 
in the euclidean partition function of a gauge theory\footnote{The spectrum of this operator is of interest for spontaneous chiral symmetry breaking in QCD because of its intimate link to the quark condensate
\cite{Banks}.}. The integration over random gauge fields can then effectively be done by
averaging over the matrix ensemble, which can be achieved by choosing a simple gaussian distribution of the matrix elements.
In the case of QCD, the random matrix partition function then describes the static modes of the Goldstone 
degrees of freedom in a finite volume at non--zero quark condensate. This represents a scaling
region (the epsilon regime or microscopic regime) where the partition 
function is fully determined by global symmetry \cite{VerUM}. 

An important research field in which a better understanding of the tools available to calculations in
non--hermitean random matrix theory
could potentially be of relevance, is in the theoretical investigation of 
finite--density QCD. The notorious sign problem 
-- the fact that the euclidean lattice action used in the probability measure for configuration sampling in lattice simulations
acquires a complex value for finite chemical potential -- effectively prevents Monte Carlo simulations in this region (see
\cite{latqcd} for reviews). In the random matrix approach,
by including a chemical potential for baryons,
the previously anti--hermitean Dirac operator acquires a hermitean piece, which renders the Dirac spectrum complex.
At present, due to the failure of lattice simulations, random matrix theory serves as an important tool of investigation
of finite--density QCD, and several analytical results have been obtained concerning the correlation functions
of ensembles with complex eigenvalues (see for example \cite{op} and references therein).
QCD at finite chemical potential has been discussed by several authors \cite{finitemureview,finitemu} and is a hot topic in random matrix theory as well as in lattice simulations.

\subsection{Analytical work on non--hermitean random matrix theory}
\label{sec-analytic}

One of the first to do analytical work on {\it non--hermitean} random matrix ensembles was Ginibre 
\cite{Gin}, who made a partially successful attempt to determine, in the mid--1960's, the eigenvalue distributions and correlation functions for three non--hermitean 
ensembles with real, complex, and real quaternion matrix elements.
His work was later completed in the orthogonal case by several authors, 
some very recently, among which Lehmann and Sommers, Edelman, Kanzieper and Akemann, Forrester and  Nagao, and Borodin and Sinclair \cite{complGin}.

After Ginibre's work, several new techniques were developed for analytically studying
various aspects of ensembles with complex eigenvalues. 
In the paper by Janik, Nowak, Papp and Zahed \cite{nhtech} 
analytical techniques for non--hermitean random matrix ensembles are reviewed.
Over the years, several interesting directions of research have been pursued, with a peak of activity around 1997.

Feinberg and Zee developed the method of hermitean reduction (hermitisation) for non--hermitean random matrix ensembles \cite{hred}.
Green's functions methods were employed by Janik, Nowak, Papp and Zahed to obtain the 
distribution of eigenvalues in the complex plane; see for example \cite{greens}.
The distribution of complex eigenvalues for symplectic ensembles of non--hermitean matrices 
was also studied by Kolesnikov and Efetov \cite{kol},
while Zabrodin has studied the growth of the domain of support of the complex eigenvalues 
with growing matrix size \cite{zab}. 

Fyodorov, Khoruzhenko and Sommers studied the cross--over from hermitean 
random matrices of Wigner--Dyson type to non--hermitean random matrices of
the type studied by Ginibre, using the technique of orthogonal polynomials \cite{cross}. 
Akemann developed the orthogonal polynomial technique for computing 
correlation functions for strong and weak hermiticity in the spirit of Fyodorov,
Khoruzhenko and Sommers, discussing also applications to the Dirac operator spectrum
in QCD with a non--zero chemical potential, see for example \cite{finitemureview};
similar studies have been performed by Kanzieper \cite{kanz}; see also the joint work on 
Ginibre's orthogonal ensemble, \cite{GinOE}.
New developments in random matrix theory (initiated by Kanzieper and further studied by Splittorff, Verbaarschot, and
others), also relevant for non--hermitean ensembles, 
include the connection between replica methods and Toda lattice hierarchies, see e.g. \cite{TL}. 

Analytical results for non--hermitean random matrix  models have been compared to lattice data in several publications. 
We cite a few (fairly randomly selected) in reference \cite{lat}.

Likely, there
will be authors that have been overlooked or forgotten in the above summary and we extend our apologies to these.

\section{Random matrix theory and symmetric spaces}
\label{sec-ss}

It is not in the scope of this article to give an introduction to random matrix theory. 
There are many good introductory texts on the subject (see for example the book by Mehta \cite{Mehta},
the book by Efetov \cite{Efe} or one of the reviews \cite{reviews}).
We will be very sketchy here and provide a minimum of detail in order to remind the reader of how the
symmetric space picture arises in the first place. For the interested, the symmetric space connection has 
been discussed in great detail for the traditional hermitean ensembles in the review \cite{SS} by Caselle and the author. 
In order not to disrupt the general discussion to review the symmetric space approach, 
we have collected all the relevant material in appendix \ref{sec-appA}. 

We reproduce here in Table \ref{tab0} the correspondence between hermitean random matrix ensembles and
symmetric coset spaces of Lie groups that was discussed and thoroughly explained in \cite{SS}. 
Let us again clarify the meaning of this table.

A random matrix ensemble modeling a Hamiltonian ${\cal H}$ is identified with the coset space 
$i{\bf P}$ of a semisimple Lie algebra ${\bf G^*}$ expressed as ${\bf G^*}={\bf K} \oplus i{\bf P}$ where ${\bf K}$ is a {\it symmetric
subgroup} (see appendix \ref{sec-appA}).  ${\bf P}$ or $i{\bf P}$ is the tangent space of a coset space of nonzero (positive or negative) curvature. This tangent space is {\it itself}
a symmetric space of zero curvature, and its group of invariance is non--semisimple because ${\bf P}$ in that
case is an abelian ideal, $[{\bf K},{\bf P}]\subset {\bf P},\ [{\bf P},{\bf P}]=0$. While all the ensembles are invariant
under the symmetry transformation $h\to khk^{-1}$ (where $k\in K$, the subgroup spanned by ${\bf K}$), the invariance of the (gaussian) random matrix partition function in this case is

\beq
\label{eq:nss}
h\to khk^{-1}+h'
\eeq

The translation invariance is what leads to the non--semisimple structure. In the table we have not explicitly
written down the coset ${\bf P}$, but in the column labeled "$X^0$" we have listed hermitean random matrix ensembles (Gaussian or Wigner--Dyson $G$, chiral~$\chi $, P--wave $P$, and Bogoliubov--DeGennes $B$) that model Hamiltonians\footnote{In the chiral ensembles we model instead a Dirac operator -- what matters is the larger symmetry group.} with different kinds of symmetries.

In the columns labeled "$G/K\ (G)$" and "$G^*/K\ (G^C/G)$" we have listed the pairs of positively and negatively
curved symmetric coset spaces that correspond to ${\bf P}$, i.e. ${\rm e}^{\bf P}$ and ${\rm e}^{i{\bf P}}$.
They are random matrix ensembles modeling scattering matrices and transfer matrices, respectively, and these
ensembles are listed with their traditional names in the columns labeled "$X^+$" and "$X^-$" ($C$ for circular,
$S$ for S--matrix and $T$ for transfer matrix ensembles). We can
view the scattering matrix of a physical system with Hamiltonian ${\cal H}$ as ${\rm e}^{i{\cal H}}$ \cite{Dyson}
(at least in a finite range).
Therefore the ensemble of the scattering matrix and the ensemble of the Hamiltonian of a physical system are listed on the same line in the table.
Since there is no such simple relationship between the Hamiltonian and the transfer matrix of a given physical system (in particular it is {\it not} ${\rm e}^{\cal H}$) the ensemble of the transfer matrix is not on the same line.
But it is nevertheless somewhere in the table in the column labeled "$X^-$". 

In the columns labeled "$m_o$", "$m_l$", "$m_s$" we find the multiplicities of the {\it restricted} ordinary, long and short roots  of the root lattice pertaining to the symmetric space. This root lattice is in many cases different
from the one inherited from the complex extension algebra ${\bf G^C}$. Note also that the total compact Lie algebra ${\bf G}$ is the sum of the algebra subspace ${\bf P}$ and the symmetric subalgebra ${\bf K}$, and the non--compact algebra ${\bf G^*}$ is obtained from it by the "Weyl unitary trick":

\beq
\label{eq:Weyl}
{\bf G}={\bf K}\oplus {\bf P},\ \ \ \ \ {\bf G^*}={\bf K}\oplus i{\bf P}
\eeq

The algebra ${\bf K}$ spans the subgroup of invariance of the random matrix ensemble. Inspection of
the table shows that for the Wigner--Dyson ensembles with $\beta=1,\ 2,\ 4$ it is $SO(N)$, $SU(N)$, and 
$USp(2N)$ -- a well--known result. We note also
that the root multiplicities $m_o$, $m_l$, $m_s$ of the ordinary, long and short roots 
of the restricted root lattice belonging to a particular symmetric space, together with the sign of the curvature 
of the corresponding manifold, fully determine
the Jacobian of the transformation to eigenvalue space in the random matrix ensemble
(equation (\ref{eq:J_j}) in appendix \ref{sec-appA}). 
The empty spaces in the table correspond to random matrix
ensembles whose physical applications have not to our knowledge been discussed in the literature so far,
but they are fully defined in the symmetric space picture.

Physical applications of each of the ensembles have been discussed in Part II of \cite{SS}, except T$^-_{4,1,4}$ that was obtained later, in \cite{nano}, as the transfer matrix ensemble of a carbon nanotube with
symplectic symmetry and an odd number of transmitting channels.
In this paper, physical observables pertaining to the carbon nanotube 
-- the conductance\footnote{Since
the particular carbon nanotubes studied in \cite{nano} have a permanently conducting channel due to an unpaired transmission eigenvalue, $\delta g$ is actually the correction to the conductance due to this channel.}
 $\delta g$, the localization length $\xi $ describing its exponential decay,
and its variance describing universal conductance fluctuations -- 
were obtained using the mapping to a symmetric space and expressed entirely in terms of group theoretical quantities. Specifically,

$$\langle {\rm ln}\,{\delta g}\rangle =-\frac{2s}{\gamma}(m_l+m_s/2)$$

$$\xi=\frac{l\gamma }{m_l+m_s/2} $$

$$\frac{{\rm Var} (\ln \, \delta g)}{\langle \ln \, \delta g \rangle}=\frac{2}{m_l+m_s/2}$$

where $s$ is the dimensionless length of the carbon nanotube expressed in terms of the mean free path $l$ of the conduction electrons, and $\gamma $ is a constant depending on the total number of eigenvalues and on the
multiplicities $m_o$, $m_l$, $m_s$ of the ordinary, long and short roots, respectively \cite{nano}.

\begin{table}
\caption{Irreducible symmetric spaces and some of their random
matrix theory realizations. Some random matrix ensembles with known
physical applications are listed in the columns labelled $X^+$, $X^0$
and $X^-$ and correspond to symmetric spaces of positive, zero and
negative curvature, respectively (for references see \cite{SS}).  Extending the notation used in the
applications of chiral random matrices in QCD, where $\nu $ is the
topological winding number, we set $\nu \equiv p-q$.  
\label{tab0}}
\vskip5mm

\hskip-1.3cm
\begin{tabular}{|l|l|l|l|l|l|l|l|l|l|}
\hline
$\begin{array}{c}Restricted\\ root\ space\end{array}$ & $\begin{array}{c}Cartan\\ class \end{array}$ & $G/K\ (G)$ & $G^*/K\ (G^C/G)$ & $m_o$ & $m_l$ & $m_s$ & $X^+$ & $X^0$ & $X^-$\\
\hline

$A_{N-1}$     & A    & $SU(N)$                 & $\frac{SL(N,C)}{SU(N)}$ & 2 & 0 & 0 & C$^+_{2,0,0}$ & G$^0_{2,0,0}$ & ${\rm T}^-_{2,0,0}$ \\
$A_{N-1}$     & AI   & $\frac{SU(N)}{SO(N)}$   & $\frac{SL(N,R)}{SO(N)}$ & 1 & 0 & 0 & C$^+_{1,0,0}$ & G$^0_{1,0,0}$ & ${\rm T}^-_{1,0,0}$\\
$A_{N-1}$     & AII  & $\frac{SU(2N)}{USp(2N)}$ & $\frac{SU^*(2N)}{USp(2N)}$ & 4 & 0 & 0 & C$^+_{4,0,0}$&G$^0_{4,0,0}$& ${\rm T}^-_{4,0,0}$\\
\hskip-2mm $\begin{array}{l} BC_q\ {\scriptstyle (p>q)} \\ C_q\  {\scriptstyle (p=q)} \end{array}$
                          & AIII & $\frac{SU(p+q)}{SU(p)\times SU(q)\times U(1)}$ & $\frac{SU(p,q)}{SU(p)\times SU(q)\times U(1)}$ & 2 & 1 & $2\nu $  & \hskip-2mm $\begin{array}{l}  \\ {\rm S}^+_{2,1,0}\end{array}$  &
 $\chi^0_{2,1,2\nu }$ &
 \hskip-2mm $\begin{array}{l}  \\ {\rm T}^-_{2,1,0} \end{array}$  \\
\hline

$B_N$          & B    & $SO(2N+1) $               & $\frac{SO(2N+1,C)}{SO(2N+1)}$ & 2 & 0 & 2  &   & P$^0_{2,0,2}$ &  \\

\hline

$C_N$ & C    & $USp(2N)$                               &$\frac{Sp(2N,C)}{USp(2N)}$ & 2 & 2 & 0 & B$^+_{2,2,0}$& B$^0_{2,2,0}$ & ${\rm T}^-_{2,2,0}$\\
$C_N$ & CI   & $\frac{USp(2N)}{SU(N)\times U(1)}$      & $\frac{Sp(2N,R)}{SU(N)\times U(1)}$& 1 & 1 & 0  & B$^+_{1,1,0}$ & B$^0_{1,1,0}$  & T$^-_{1,1,0}$  \\
\hskip-2mm $\begin{array}{l} BC_q\  {\scriptstyle (p>q)} \\  C_q\  {\scriptstyle (p=q)} \end{array}$ & CII & $\frac{USp(2p+2q)}{USp(2p)\times USp(2q)}$ & $\frac{USp(2p,2q)}{USp(2p)\times USp(2q)}$& 4 & 3 & $4\nu $ &   & $\chi^0_{4,3,4\nu }$ & \hskip-2mm $\begin{array}{l} \\ {\rm T}^-_{4,3,0}\end{array}$   \\
\hline

$D_N$    & D          &   $SO(2N)$                            & $\frac{SO(2N,C)}{SO(2N)}$ & 2 & 0 & 0 & B$^+_{2,0,0}$  &  B$^0_{2,0,0}$  & ${\rm T}^-_{2,0,0}$\\
$C_N$    & DIII-e  & $\frac{SO(4N)}{SU(2N)\times U(1)}$     & $\frac{SO^*(4N)}{SU(2N)\times U(1)}$ & 4 & 1 & 0 & B$^+_{4,1,0}$ & B$^0_{4,1,0}$ & T$^-_{4,1,0}$  \\
$BC_N$   & DIII-o  & $\frac{SO(4N+2)}{SU(2N+1)\times U(1)}$ & $\frac{SO^*(4N+2)}{SU(2N+1)\times U(1)}$& 4 & 1 & 4 &  & P$^0_{4,1,4}$ & T$^-_{4,1,4}$  \\
\hline

\hskip-2mm $\begin{array}{l} B_q\ {\scriptstyle (p>q)}\\ D_q\ {\scriptstyle (p=q)} \end{array}$
            & BDI & $\frac{SO(p+q)}{SO(p)\times SO(q)}$   & $\frac{SO(p,q)}{SO(p)\times SO(q)}$ & 1 & 0 & $\nu $ &   & $\chi^0_{1,0,\nu }$ & $\begin{array}{l} \\ {\rm T}^-_{1,0,0} \end{array}$\\

\hline
\end{tabular}
\end{table}

\section{Symmetries defining random matrix ensembles}
\label{sec-sym}

\subsection{Summary of results}
\label{sec-summary}

In this paper we take a first step towards the identification of random {\it non--hermitean} matrix ensembles
with symmetric spaces. Clearly, the hermitean ensembles exhaust the Cartan classification table of
symmetric spaces parametrized in terms of {\it real} coordinates. We will go beyond this classification in 
listing a number of non--hermitean
matrix ensembles defined by certain symmetry criteria. The main result of this work is to show that 
also these ensembles define symmetric spaces, but these new spaces will be complex manifolds,
in line with the fact that the corresponding sets of random matrices have complex eigenvalues.

We give explicit representations in Tables \ref{tab1}--\ref{tab5} of a large number of random matrix ensembles.
The latter are identified with the set ${\bf P}$ in the tables, whereas the symmetric subgroup in each case 
is spanned by ${\bf K}$. The algebra subspace ${\bf P}$ and the subalgebra ${\bf K}$ satisfy the commutation
relations $[{\bf K},{\bf K}]\subset {\bf K},\ [{\bf K},{\bf P}]\subset {\bf P},\ [{\bf P},{\bf P}]\subset {\bf K}$
and their sum spans the total algebra ${\bf G}$ or ${\bf G^*}$  in the notation of appendix \ref{sec-appA} and equation (\ref{eq:Weyl}).
 
The ensembles will be defined according to certain physically motivated symmetry criteria described in detail below.
Among the ensembles listed in Tables \ref{tab1}--\ref{tab5} we find as special cases the ten hermitean ensembles
of Table \ref{tab0} corresponding to an even number of eigenvalues (this excludes Cartan classes B and DIII--odd)  
and the three Ginibre ensembles. We choose for simplicity even--dimensional matrices and we fix the chiral
symmetry operator to have an equal number of positive and negative eigenvalues. 

Let us now describe the
symmetry criteria we have used to obtain the explicit representations of the ensembles.
 
\subsection{The symmetry criteria} 
\label{sec-symcrit} 

We use as a basis for our classification the symmetry criteria defined in the paper by Bernard and LeClair,
\cite{BLC}. The authors of this paper provide a classification of non--hermitean random matrices
in terms of these symmetries. However, their paper does not provide explicit matrix representations for the
ensembles of random matrices (they do, however, list inequivalent representations of the matrices defining
the symmetries). Though not very explicit in their paper\footnote{On page 3 they do, however, refer to the operator
${\cal H}$ as a "doubled Hamiltonian", which shows that this is what they had in mind.}, what they classify is cosets of an algebra, but they are referred to loosely as "groups" throughout the paper. In this paper we give explicit
representations of all these matrix ensembles and show that all the ensembles classified in \cite{BLC} represent
algebra subspaces of the type ${\bf P}$ in equation (\ref{eq:commrel}) in the Appendix, reproduced below 
for convenience. The remaining algebra subspace is a symmetric subalgebra ${\bf K}$. 
As already mentioned, the two algebra subspaces satisfy the commutation relations

\beq
\label{eq:commrel'}
[{\bf K},{\bf K}]\subset {\bf K},\ \  [{\bf K},{\bf P}]\subset {\bf P},\ \
[{\bf P},{\bf P}]\subset {\bf K}
\eeq

defining ${\bf K}$ as a symmetric subalgebra. 

We will now, following \cite{BLC},  define the discrete symmetries we used in defining the cosets 
${\bf P}$. We also keep the notation of this reference and reproduce the definitions used by Bernard and LeClair here. It is desirable to limit the number of symmetry classes under consideration by some physically motivated conditions. The resulting classification will be complete in the sense that all the ensembles defined by the symmetry criteria to be defined below will be classified in tables \ref{tab1} through \ref{tab5}. 
Following \cite{BLC}
we consider unitary transformation matrices $p$, $c$, $q$ and $k$ such that a complex matrix $h$ is transformed
according to

\bea
\label{eq:symtransf}
h=-php^{-1},\ \ \ \ \ h=\epsilon_c ch^Tc^{-1},\ \ \ \ \ h=qh^\dagger q^{-1},\ \ \ \ \ h=kh^*k^{-1}
\eea
\bea
pp^\dagger =
cc^\dagger =
qq^\dagger =
kk^\dagger =1
\eea

where $\epsilon_c =\pm $. Introducing similar signs in the relations for $q$ and $k$ symmetries does not lead to
new ensembles, because they can be obtained by the substitution $h\to ih$ \cite{BLC}. The above symmetries will be referred to as P--type, C--type, Q--type and K--type symmetry, respectively.

The physical motivation for these symmetries is as follows. 

The P--symmetry is chiral symmetry $\{ h,p\} =0$ with $p$ playing the role of 
$\gamma_5$. Requiring that the p--action be of order two (i.e. that by applying $p$ twice we 
get back the same operator $h$) implies $p^2=1$. Following \cite{BLC}, we will
require that all the symmetries are of order two.

The C--type symmetries are the usual anti--unitary symmetries for hermitean operators. 
For $\epsilon_c =+$ and hermitean $h$ it is equivalent to time reversal symmetry, 
$[h,cK]=0$ where $c$ is unitary and $K$ denotes complex conjugation. Since
$cc^*=\pm 1$,\footnote{Time reversal invariance arises from the fact that
(in the absence of magnetic fields), if $\Psi({\bf r},t)$ is a
solution of the Schr\"odinger equation, so is $\Psi^*({\bf r},-t)$. We
define the action of the (anti--unitary) time-reversal operator $T$ on
a state $\Psi$ by $T\Psi=c\Psi^*$, where $c$ is a unitary
operator. $T$ should reverse the sign of spin and total angular
momentum. This requirement can be satisfied by the choice $c={\rm
e}^{i\pi S_y}$ for spin rotation or $c={\rm e}^{i\pi J_y}$ for space
rotation, with a standard representation of spin or space rotation
matrices. Now $T^2=cc^*={\rm e}^{2i\pi S_y}=(-)^n=\pm 1$. These two
cases correspond to integer and half--odd integer spin (or presence or
absence of space rotation invariance in case of $J_y$). } it follows $c^Tc^{-1}=\pm 1$. 
For $\epsilon_c=-$ and hermitean $h$, the C--type symmetry is equivalent to particle--hole
symmetry, $\{ h,cK\}=0$ (see for example \cite{AZ} where this symmetry is stated in
the form $X\equiv i{\cal H} = -\Sigma_xX^T\Sigma_x$, $\Sigma_x\equiv \sigma_x\otimes 1\!\!1$
for the Bogoliubov--De Gennes Hamiltonian ${\cal H}$).

The physical requirement that $c^Tc^{-1}=\pm 1$ can be obtained by requiring that
the action of $c$ is of order two:

\beq
h=\epsilon_c c(\epsilon_c ch^Tc^{-1})^Tc^{-1}
\eeq

This implies $c^{-1}c^T=\pm 1$, which corresponds to choosing $c$ symmetric or antisymmetric, respectively.
Note that these two cases correspond in the classical ensembles to presence or absence of space--rotation invariance or alternatively, integer or half--odd integer spin.

The Q--type symmetry reduces to the statement that the ensemble is hermitean if $q=1\!\!1$. Therefore  we
identify all such ensembles with one of the Cartan classes of symmetric spaces. Note 
also that for hermitean ensembles this symmetry becomes trivial, and so it is only meaningful
for non--hermitean ensembles, where it imposes conditions of (anti--)hermiticity on the random matrix 
or on its sub--blocks. 
The transformations are unitary, so it follows by applying the same transformation twice that
$q^{-1}q^\dagger=+1$ (in this case we get, because of unitarity, $q^2=1$).  

For hermitean matrices the K--type symmetry  is redundant -- it is the same as time--reversal symmetry --
but for non--hermitean ensembles it imposes different kinds of reality conditions on the ensembles. 
From the requirement that the symmetry be of order two we obtain $kk^*=\pm 1$, which implies that $k$
is symmetric or antisymmetric, respectively.

In \cite{BLC}, if two matrices $h$ were related by a unitary similarity transformation, they were considered
to be equivalent. We will keep this rule, which is at the heart of the coset space structure, but we will report in the comments to the entries in Table \ref{tab1} through Table \ref{tab5} 
and in tables \ref{tab-equi20} and \ref{tab-equi2425} a few equivalent forms of some of the 
ensembles (all related by unitary transformations), for the following reason. As we compared our tables to
some ensembles used in the literature, we noticed that sometimes a unitarily equivalent form 
of an ensemble was used, which makes its identification in the tables more difficult. 
We therefore included 
some of the obviously equivalent forms corresponding to symmetry implementations compatible with the commutativity constraints
(see below). Nevertheless, 
the identification of ensembles may not be immediate. For the PQC--symmetric
ensembles however, the equivalence relations between different ensembles are transferred to an equivalence relation for the sub--blocks of $q$ and $c$. Therefore, some of the equivalent ensembles can be read off from the
corresponding equivalence relations for non--chiral ensembles and are not explicitly reported.

According to the above, we consider $h$ and $h'$ equivalent whenever

\beq
h'=uhu^\dagger
\eeq

with $u$ unitary. This translates into similar equivalence relations for the symmetry implementation 
matrices $p$, $c$, $q$ and $k$ defining the ensemble \cite{BLC}:

\beq
\label{eq:simrels}
p'=upu^\dagger ,\ \ \ \ \ c'=ucu^T,\ \ \ \ \ q'=uqu^\dagger ,\ \ \ \ \ k'=uku^T 
\eeq

where we have used the fact that $uu^\dagger = u^\dagger u=1$.
Therefore, two ensembles in the tables characterized by a subset of $\{ p,c,q,k\}$ and $\{ p',c',q',k'\}$
respectively, are unitarily equivalent if the subsets are related by equation (\ref{eq:simrels}).

In case of just one symmetry, we consider the unitarity of $p,\ c,\ q,\ k$ together with the constraints
$p^2=1$, $cc^*=\pm 1$, $q^2=1$, $kk^*=\pm 1$ to obtain all the possible inequivalent forms of
these matrices. Let us first consider the matrix $p$. It satisfies 
$pp^\dagger=p^\dagger p=p^2=1$, which means that it is diagonalizable with eigenvalues $\pm 1$.
The solution $p=1$ implies $h=0$. We will, like in \cite{BLC}, for simplicity consider only the case of an equal number of $+1$ and $-1$ eigenvalues\footnote{This choice has no bearing on the symmetric space structure. That is, for an unequal number of positive and negative eigenvalues, we obtain similar symmetric spaces.}, not least for the purpose of limiting the number of ensembles,
and we fix $p={\rm diag}(1,...,1,-1,...,-1)$.
This means that all our
chirally invariant ensembles will have the usual block--structure.
$c$ and $k$ can be symmetric or antisymmetric, while the constraint on $q$ excludes an antisymmetric $q$. The possibilities are listed 
in Table \ref{tab1}. Note that for ensembles {\bf 8} and {\bf 15} we have chosen a form for $k$ that displays the real
quaternion structure of the matrix elements (see the specific comment in subsection \ref{sec-spec-com}
and appendix \ref{sec-appB}).

If two or more symmetries are present, we will demand, following \cite{BLC}, that the symmetries
commute. For example, if we have PC symmetry, we demand $-p(\epsilon_cch^Tc^{-1})p^{-1}=
\epsilon_cc(-php^{-1})^Tc^{-1}$. For all the pairs of symmetries we then obtain the commutativity 
constraints of \cite{BLC}, we reproduce below:

\bea
\label{eq:commcon}
c=\pm pcp^T \equiv \epsilon_{cp}pcp^T,\ \ \ \ \ p=\pm kp^*k^\dagger ,\nonumber \\
q=\pm pqp^\dagger \equiv \epsilon_{pq} pqp^\dagger ,  q=\pm cq^*c^\dagger \equiv \epsilon_{cq} cq^*c^\dagger ,\\ 
q=\pm kq^*k^\dagger ,\ \ \ \ \ c=\pm kc^*k^T\nonumber 
\eea

(here $\epsilon_{cp},\ \epsilon_{pq}$ and $\epsilon_{cq}$ are signs).
The constraints (\ref{eq:commcon}) can be used to determine all the non--equivalent forms of 
the symmetry implementation matrices for any subset of symmetries.  
Starting from $p$,

\beq
p=\left(\begin{array}{cc} 1\!\!1 & 0 \cr
                                             0  & -1\!\!1\end{array}\right)
\eeq

we can then determine the possible forms of $c$, $q$ and $k$ compatible with eqs.~(\ref{eq:commcon}).
As an example of this procedure, let's consider an ensemble with PC symmetry.
For $p=p^T$, (\ref{eq:commcon}) gives either $[p,c]=0$ or $\{ p,c\}=0$. The former leads to a block--diagonal 
$c$ and the latter to a block--offdiagonal $c$. Also, $c$ is always either symmetric or antisymmetric. Inspection shows that all the antisymmetric
forms 

\beq
c=\left(\begin{array}{cc} 0 & 1\!\!1 \cr -1\!\!1 & 0 \end{array}\right)
\simeq \left(\begin{array}{ccccc} 0 &1&&& \cr -1 & 0 &&& \cr && \ddots && \cr &&& 0&1\cr &&& -1&0 \end{array}\right)
\simeq \left(\begin{array}{cccc} &&0&1\!\!1 \cr&&-1\!\!1 &0 \cr 0& 1\!\!1&&\cr -1\!\!1&0 \end{array}\right)
\simeq \left(\begin{array}{cccc} 0&1\!\!1 &&\cr -1\!\!1 &0 && \cr & &0&1\!\!1 \cr && -1\!\!1 &0 \end{array}\right)
\eeq

are related by unitary transformations. This does {\it not} mean that only one of them has to be 
considered in a general ensemble, as the presence of other symmetries may prevent the ensemble
from being equivalent to another ensemble with the same type of symmetries, because also these
other symmetries have to transform according to (\ref{eq:simrels}) with the {\it same} unitary transformation
that relates $c$ to $c'$. Thus, different equivalent forms for $c$ leave open the possibility of the 
corresponding ensembles being equivalent, but only if the other symmetries of the ensembles 
are transformed into each other by the same transformation. 
There are also three equivalent possible symmetric forms of $c$

\beq
c=\left(\begin{array}{cc}1\!\!1 & 0\cr 0&1\!\!1\end{array}\right)\simeq 
\left(\begin{array}{cc}1\!\!1 & 0\cr 0&-1\!\!1\end{array}\right)\simeq 
\left(\begin{array}{cc}0&1\!\!1\cr 1\!\!1 & 0 \end{array}\right)
\eeq

In the same way we find two equivalence classes for $k$ which are identical to the ones above. For $q$ we find
the two equivalence classes

\beq 
q=1\!\!1\ \ \ \ \ {\rm or}\ \ \ \ \ q\in \left\{ \pzzm , \zppz \right\}
\eeq

For PC symmetry, it turns out that the two inequivalent forms

\beq
c=\left(\begin{array}{cc} 0 & 1\!\!1 \cr -1\!\!1 & 0 \end{array}\right),\ \ \ \ \ c'=\left(\begin{array}{cc}0&1\!\!1\cr 1\!\!1 & 0 \end{array}\right) 
\eeq 

give rise to the same ensembles, but with opposite sign of $\epsilon_c$ in the two cases (table entry {\bf 11}).
Entries {\bf 10} and {\bf 11} have unitarily equivalent forms of $c$, but $p$ does not transform into itself under any unitary transformation taking the two different antisymmetric forms of $c$ into each other, so they represent two inequivalent ensembles. 

To summarize, for PC symmetry we can have either a block--diagonal or a block--offdiagonal $c$, and in addition $c^T=\pm c$ which would give rise to four distinct ensembles, but because the inequivalent forms in the equation above happen
to give rise to the same ensembles,
for PC there are only three different ensembles instead of four, and they are listed in Table~\ref{tab2}. 

It is also noteworthy that the classified symmetry combinations exhaust all the symmetry classes, as
having more symmetries of the same type does not lead to anything new. For example,
two C--type symmetries generate a P--type symmetry if the product of their $\epsilon_c$'s is $-1$, etc. 
Also, combinations of the type QK or CK leads to one of the already listed classes, since two of the symmetries C, Q,  and K combine to give a third type of symmetry. We
refer the reader to \cite{BLC} for more discussion on this point.

Having now explained through an example how ensembles are distinguished, we consider this issue clear
and we will not go through the details of all the classification. We refer to Tables \ref{tab1} through \ref{tab5}
for a listing of all inequivalent ensembles.  

The integration measures in the random matrix partition functions corresponding to the ensembles will not be discussed in detail here --
they should be defined by restriction of the invariant group measure to the respective cosets. A gaussian
invariant measure is the simplest choice and is usually sufficient, since choosing a more complicated
polynomial integration measure often does not change the universal properties of correlation functions.
We expect an intimate relationship might exist between the non--hermitean Jacobians and restricted root lattices, like for
hermitean ensembles.
We will also not even make any attempt here at identifying the new ensembles with coset spaces of groups, so in the last
column of Tables \ref{tab1} through \ref{tab5} we therefore indicate only where to find the 
ensembles known from the classifications of Cartan and Ginibre and leave the other spaces empty, indicating
that the particular ensemble is of a new, non--hermitean type.

\section{The matrix ensembles}
\label{sec-com}

\subsection{General comments on the tables}
\label{sec-gen-com}

In Tables \ref{tab1} through \ref{tab5} we list in the first column the types of symmetry defining the ensemble
of random (non--)hermitean matrices. All the entries marked with an asterisk in this column are commented 
on in subsection \ref{sec-spec-com}, and the lines are numbered for easy reference. 

In the second column we list the explicit implementation of the matrices $p$, $c$, $q$ and $k$ defining the symmetries in accordance
with equations (\ref{eq:symtransf}) and (\ref{eq:commcon}). Often two alternatives are given. Whenever there are two sign alternatives on the same line
of this column, the two alternative signs in the following columns correspond to these two alternatives in the same order. 

The combinations of symmetries are constrained by the commutativity constraints 
(\ref{eq:commcon}) in case of two or more symmetries. 
In case a C--symmetry is present, the sign of $\epsilon_c$ defines whether
it is a time--reversal type symmetry ($\epsilon_c=+$) or a particle--hole type symmetry ($\epsilon_c=-$).

In some cases there are other implementations of C--symmetry that are possible (and compatible
with commutativity constraints) than the ones given in the tables. For example, both for ensembles {\bf 4} and {\bf 10},
a possible representation for $c$ (compatible with the commutativity constraint for $p$ and $c$ for the 
ensemble with both P and C symmetry) is

{\small $$ c=\isy $$}

This structure of $c$ (or $k$) is the one that leads to a representation in terms of {\it real quaternions} in the ensembles
having this possibility of choice of basis. 
However, such a $c$ leads to a matrix representation in terms of {\it complex} quaternions in the ensembles that
do {\it not} have this possibility.
Since the latter do not form a field, we prefer the
form of $c$ given in the table. We will follow this rule throughout the tables whenever there is an ambiguity of this kind.

As we have seen, there is some margin of flexibility and ambiguity in the implementation
of symmetries due to the possibility of unitarily equivalent representations. We will to a large extent follow reference \cite{BLC} in the symmetry classification, but in certain cases we deviate from it where we
find this motivated (in particular, when we want to display the real quaternion structure of an ensemble known
to possess it, we may choose to display a unitarily equivalent ensemble to the one listed in \cite{BLC}). This does not
of course change any of its symmetry properties. It should also be mentioned that we have two more ensembles with respect to reference \cite{BLC}, ensembles {\bf 29} and {\bf 30}.
Probably these were just forgotten in \cite{BLC} -- the corresponding combination of signs of 
$\epsilon_{cp},\ \epsilon_{pq},\ \epsilon_{cq}$ ($++-$) was included on line 2 in the list of equivalence relations at the bottom of p. 5 of the electronic version of \cite{BLC}, but apparently the corresponding symmetry class was not listed.
Also, our tables are more redundant than the listing in \cite{BLC} as we include two sign alternatives on almost every
line, some of which are unitarily equivalent.

In the columns labeled "${\bf P}$" and "${\bf K}$" we list the explicit form of random matrix ensembles with the
given symmetries and the symmetric subalgebra defining the subgroup of invariance $K$,

\beq
h\to khk^{-1}\ \ \ \ \ \ \ (k\in K)
\eeq

of the ensemble\footnote{As explained before, for symmetric spaces of zero curvature, i.e. algebra subspaces, the subgroup of invariance is actually larger and given by (\ref{eq:nss}).}. 
The random matrix ensemble is always identified with the algebra subspace ${\bf P}$,
while the symmetric subalgebra ${\bf K}$ obtained from eq. (\ref{eq:commrel}) spans the subgroup of
invariance. 
 Note that we do not distinguish between subspaces ${\bf P}$ and $i{\bf P}$ in our tables, as they are
equivalent, and both of them are reported in the same column. 
We stress that the column "${\bf P}$" contains algebra cosets corresponding to
random {\it Hamiltonians} ${\cal H}$ or $i{\cal H}$ (or to Dirac operators in case of chiral ensembles) -- not scattering or transfer matrices. The latter should be obtained by 
exponentiating ${\bf P}$ or $i{\bf P}$ to obtain symmetric coset spaces of Lie groups (see  section \ref{sec-ss} or \cite{SS}). 

In case the column corresponding to ${\bf K} $ is empty, or the corresponding entry is identical to the one in
the column for ${\bf P}$ -- possibly after using the Weyl unitary trick ${\bf P}\to i{\bf P}$ -- it means that ${\bf P}$ is an algebra, i.e. $[{\bf P},{\bf P}]\in {\bf P}$. This happens whenever
the symmetric space is a group manifold  and not just a coset space, for example for Cartan classes A, C, D (B is not included because we limited ourselves to even--dimensional representations). 

We observe that in tables \ref{tab1} through \ref{tab5}, ensembles corresponding to opposite values of $\epsilon_c$
correspond to the same symmetric subalgebra and are in this sense dual to each other. We find that opposite values 
of $\epsilon_c$ correspond to
random matrix ensembles ${\bf P}$ with differing structure; however such ensembles are considered 
equivalent even though not related simply by the Weyl unitary trick \cite{BLC}, \cite{Gilmore} (Ch. 9). 

Just like for the hermitean ensembles, the algebra subspace ${\bf P}$ is the tangent space 
of a symmetric space. The difference is that these new symmetric spaces are {\it complex} manifolds
(i.e. when the random matrix ensemble is non--hermitean), which can be seen as an even--dimensional real manifold on the tangent space of which we have defined a {\it complex structure}. The discussion of these complex
symmetric spaces will be deferred to a future publication. We know they are symmetric because ${\bf K}$
and ${\bf P}$ satisfy the commutation relations (\ref{eq:commrel}) defining a symmetric space. This has been
checked for each ensemble in the process of computing the explicit forms of the algebra subspaces from the defining symmetry relations.

If ${\bf P}$ is a hermitean Cartan class or one of the three Ginibre ensembles, the corresponding class is indicated in the last column labeled "Symmetry class" where "Gin1" means it is the Ginibre ensemble with $\beta =1$ etc.
The Ginibre ensembles have  at most one symmetry and are all found in the first table.

In appendix \ref{sec-appB} we have collected the relevant properties of quaternions that are needed for the
quaternion real ensembles. 

{\small
\begin{table}[ht]
\hskip-1.7cm
\caption{Explicit representations of random matrix ensembles having no symmetries or only one symmetry. 
See sections \ref{sec-sym} and \ref{sec-com} for explanations and comments. In the last column
of tables \ref{tab1} through \ref{tab5} the Cartan classes known from the classification of hermitean
random matrix ensembles have been indicated, as well as the three Ginibre ensembles.
\label{tab1}}
\vskip5mm

\hskip-.7cm
\begin{tabular}{c|c|c|c|c}
\cr \hline 
{\rm Sym.}&
{\rm Implementation} &
{\bf P} &
{\bf K} &
$ \barc {\rm Symmetry}\cr  {\rm class} \ear $ \cr \hline 
{\bf 1} -- &  & A &  & Gin2\cr \hline 
{\bf 2} P & $p=\pzzm $  & $ \left( \begin{array}{cc} 0 & A \cr B & 0 \end{array} \right) $ & 
$ \left( \begin{array}{cc} X & 0 \cr 0 & Y \end{array} \right) $ & \cr \hline 
{\bf 3} C & $c=1\!\!1,\ \epsilon_c=\pm$ & 
$ \barc A \cr A=\pm A^T  \ear $   & $ \barc X \cr X=-X^T \ear $ & \cr \hline 
{\bf 4} C & $c=\zpmz $,\  $ \epsilon_c=\pm $ & $ \barc \left( \begin{array}{cc} A & B \cr C & \pm A^T \end{array} \right)  \cr B=\mp B^T,\ C=\mp C^T  \ear $ & $ \barc \left( \begin{array}{cc} X & Y \cr Z & -X^T \end{array} \right) \cr Y=Y^T,\ Z=Z^T \ear $  & \cr \hline 
{\bf 5} Q & $q=1\!\!1$ & $\barc A \cr A=A^\dagger \ear $ & & A \cr \hline 
{\bf 6} Q & $q=\pzzm $ & $ \barc \left( \begin{array}{cc} A & B \cr -B^\dagger & D \end{array} \right)  \cr A=A^\dagger,\  
D=D^\dagger\ear $ & 
$ \barc \left( \begin{array}{cc} X & Y \cr Y^\dagger  & Z \end{array} \right) \cr X=-X^\dagger ,\ Z=-Z^\dagger \ear $  & \cr \hline 
{\bf 7} K & $k=1\!\!1  $ &$\barc A \cr A=A^* \ear $ & & Gin1 \cr \hline 
{\bf 8} K$^*$ & $k=\isy $ & $ \barc A \cr A\ {\rm quaternion\ real} \cr (A^\dagger =\bar{A})\ear $  & & Gin4 \cr \hline 
\end{tabular}
\end{table}}

{\small
\begin{table}[ht]
\hskip-1.7cm
\caption{Explicit representations of random matrix ensembles with only PC symmetry.
See sections \ref{sec-sym} and \ref{sec-com} for explanations and comments.
\label{tab2}}
\vskip5mm

\hskip-1.7cm
\begin{tabular}{c|c|c|c|c}
\cr \hline 
{\rm Sym.}&
{\rm Implementation} &
{\bf P} &
{\bf K} &
$ \barc {\rm Sym.}\cr  {\rm class} \ear $ \cr \hline 
{\bf 9} PC & $\barc p=\pzzm \cr c=1\!\!1,\ \epsilon_c=\pm \ear $ & $ \left( \begin{array}{cc} 0 & A \cr \pm A^T & 0 \end{array} \right) $ & $ \barc \left( \begin{array}{cc} X & 0 \cr 0 & Y \end{array} \right) \cr
X=-X^T,\ Y=-Y^T \ear $ & \cr \hline 
{\bf10} PC & $\barc p=\pzzm \cr c=\ISY \cr \epsilon_c=\pm  \ear $ &
$ \left( \begin{array}{cc}  \!\!\! \!\!\! & \!\!\! \begin{array}{cc} A & B \cr C & D \end{array} \!\!\! \cr   
\!\!\! \begin{array}{cc} \pm D^T & \mp B^T \cr \mp C^T & \pm A^T  \end{array} \!\!\! &\!\!\! \!\!\!  \end{array} \right) $ &
$ \barc \left( \begin{array}{cc} \!\!\! \begin{array}{cc} X & Y \cr Z & -X^T \end{array} \!\!\! & \!\!\! \!\!\! \cr & \!\!\!
\begin{array}{cc} U & V \cr W & -U^T \end{array}\!\!\!  \end{array} \right)  \cr Y=Y^T,\ Z=Z^T,\cr V=V^T,\ W=W^T  \ear $ \cr \hline 
{\bf 11} PC$^*$ & $\barc p=\pzzm  \cr c=\zppz ,\ \epsilon_c=\pm \cr
{\rm or }\  c=\zpmz ,\ \epsilon_c=\mp \ear $ & $ \barc \left( \begin{array}{cc} 0 & A  \cr B & 0 \end{array} \right) , \cr A=\pm A^T, \ B=\pm B^T  \ear $ &  
$ \left( \begin{array}{cc} X & 0 \cr 0 & -X^T \end{array} \right) $ & \cr \hline 
\end{tabular}
\end{table}}

{\small
\begin{table}[ht]
\hskip-1.7cm
\caption{Explicit representations of random matrix ensembles with only PQ symmetry.
See sections \ref{sec-sym} and \ref{sec-com} for explanations and comments.
\label{tab2'}}
\vskip5mm

\hskip-.2cm
\begin{tabular}{c|c|c|c|c}
\cr \hline 
{\rm Sym.}&
{\rm Implementation} &
{\bf P} &
{\bf K} &
$ \barc {\rm Symmetry}\cr  {\rm class} \ear $ \cr \hline 
{\bf 12} PQ$^*$ & $\barc p=\pzzm \cr q=\pzzpm \ear $ & $ \left( \begin{array}{cc} 0 & A \cr \pm A^\dagger & 0 \end{array} \right) $ & 
$ \barc  \left( \begin{array}{cc} X & 0 \cr 0 & Y \end{array} \right)  \cr X=-X^\dagger ,\ Y=-Y^\dagger  \ear $& $\barc {\rm AIII}\cr 
{\rm (for}\ q=1\!\!1{\rm )} \ear $ \cr \hline 
{\bf 13} PQ & $\barc p=\pzzm \cr q=\zppz \ear $ &$ \barc  \left( \begin{array}{cc}  0 & A \cr B & 0 \end{array} \right) \cr 
A=A^\dagger,\ B=B^\dagger  \ear $ & $ \left( \begin{array}{cc}  X & 0 \cr 0 & -X^\dagger \end{array} \right) $ & \cr \hline 
\end{tabular}
\end{table}}

{\small
\begin{table}[ht]
\hskip-1.7cm
\caption{Explicit representations of random matrix ensembles with only PK symmetry.
See sections \ref{sec-sym} and \ref{sec-com} for explanations and comments.
\label{tab2''}}
\vskip5mm

\hskip-1.4cm
\begin{tabular}{c|c|c|c|c}
\cr \hline 
{\rm Sym.}&
{\rm Implementation} &
{\bf P} &
{\bf K} &
$ \barc {\rm Symmetry}\cr  {\rm class} \ear $ \cr \hline 
{\bf 14} PK$^*$ & $\barc  p=\pzzm  \cr k=\pzzpm \ear $ & $ \barc  \left( \begin{array}{cc} 0 & A \cr B & 0 \end{array} \right)  \cr A=\pm A^*,\ B=\pm B^*  \ear $ & $ \barc  \left( \begin{array}{cc} X & 0 \cr 0 & Y \end{array} \right)  \cr X=X^*,\ Y=Y^* \ear $ & \cr \hline 
{\bf 15} PK$^*$ & $\barc  p=\pzzm  \cr k=\isy \ear $ & $ \barc  \left( \begin{array}{cc} 0 & A \cr B & 0 \end{array} \right) \cr  A, B\  {\rm quaternion\ real}  \ear $ & $ \barc \left( \begin{array}{cc} X & 0 \cr 0 & Y \end{array} \right) \cr  X,Y\  {\rm quaternion\ real}  \ear $ & \cr \hline 
{\bf 16} PK$^*$ & $\barc p=\pzzm \cr k=\zppmz \ear $ & $ \left( \begin{array}{cc} 0 & A \cr \pm A^* & 0 \end{array} \right) $  & $ \left( \begin{array}{cc} X & 0 \cr 0 &  X^* \end{array} \right) $  & \cr \hline 
\end{tabular}
\end{table}}

{\small
\begin{table}[ht]
\hskip-1.7cm
\caption{Explicit representations of random matrix ensembles with only QC symmetry (continued in Table \ref{tab3'}).
See sections \ref{sec-sym} and \ref{sec-com} for explanations and comments.
\label{tab3}}
\vskip5mm

\hskip-1.2cm
\begin{tabular}{c|c|c|c|c}
\cr \hline 
{\rm Sym.}&
{\rm Implementation} &
{\bf P} &
{\bf K} &
$ \barc {\rm Symmetry}\cr  {\rm class} \ear $ \cr \hline 
{\bf 17} QC & $\barc q=1\!\!1 \cr c=1\!\!1,\ \epsilon_c=\pm \ear $ & $ \barc  A \cr A= \pm A^*= \pm A^T \ear $   & $ \barc  X \cr X=X^*=-X^T \ear $  & $ \barc {\rm AI} \cr {\rm D} \ear $ \cr \hline 
{\bf 18a} QC$^*$ & $ \barc q=1\!\!1 \cr c=\isy , \cr \epsilon_c=+ \ear $ & $ \barc  A \cr A=\bar{A};\cr  A\ {\rm quaternion\ real} \cr
(A^\dagger=\bar{A}) \ear $   & $  \barc X\cr X=-\bar{X} \cr X\ {\rm quaternion\ real} \cr (X^\dagger =\bar{X}) \ear $  & AII\cr \hline 
{\bf 18b} QC$^*$& $ \barc q=1\!\!1 \cr c=\zppmz , \cr \epsilon_c=- \ear $ & $ \barc  \left( \begin{array}{cc} A & B \cr \mp B^* & -A^* \end{array} \right)  \cr A=A^\dagger,\ B=\mp B^T  \ear $   & $ \barc  \left( \begin{array}{cc} X & Y \cr -Y^\dagger & -X^T \end{array} \right)  \cr  X=-X^\dagger,\ Y=Y^T \ear $ & $\barc {\rm (D)} \cr {\rm C} \ear $ \cr \hline 
{\bf 19} QC$^*$ & $\barc q=\pzzm \cr c=1\!\!1,\ \epsilon_c=\pm \ear $ & $ \barc  \left( \begin{array}{cc} A & B \cr \pm B^T & C \end{array} \right)  \cr A=\pm A^*=\pm A^T,\cr B=\mp B^*,\cr C=\pm C^* = \pm C^T  \ear $  & $ \barc  \left( \begin{array}{cc} X & Y  \cr Y^\dagger & Z \end{array} \right)  \cr X=X^*=-X^T,\cr Y=-Y^*,\cr Z=Z^*=-Z^T \ear $ & \cr \hline 

\end{tabular}
\end{table}}

{\small
\begin{table}[ht]
\hskip-1.7cm
\caption{Explicit representations of random matrix ensembles with only QC symmetry (continued from Table \ref{tab3}).
See sections \ref{sec-sym} and \ref{sec-com} for explanations and comments.
\label{tab3'}}
\vskip5mm

\hskip-1.6cm
\begin{tabular}{c|c|c|c|c}
\cr \hline 
{\rm Sym.}&
{\rm Implementation} &
{\bf P} &
{\bf K} &
$ \barc {\rm Sym.}\cr  {\rm class} \ear $ \cr \hline 
{\bf 20a} QC$^*$ & $ \barc q=\pzzm \cr c=\isy ,\cr \epsilon_c=+  \ear $ & $ \barc  \left( \begin{array}{cc} A & iB \cr i\bar{B} & C \end{array} \right) \cr A=\bar{A},\ C=\bar{C},\cr  A, B, C\  {\rm quaternion\ real}  \ear $  & 
$ \barc  \left( \begin{array}{cc} X & iY \cr -i\bar{Y} & Z \end{array} \right)  \cr X=-\bar{X},\ Z=-\bar{Z} \cr X, Y, Z\ {\rm quaternion\ real} \ear $  & \cr \hline 
{\bf 20b} QC$^*$ & $ \barc q=\pzzm \cr c=\isy ,\cr \epsilon_c=-  \ear $ & $ \barc  \left( \begin{array}{cc} iA & B \cr -\bar{B} & iC \end{array} \right) \cr A=-\bar{A},\ C=-\bar{C},\cr  A, B, C\  {\rm quaternion\ real}  \ear $  & 
$ \barc  \left( \begin{array}{cc} X & iY \cr -i\bar{Y} & Z \end{array} \right)  \cr X=-\bar{X},\ Z=-\bar{Z} \cr X, Y, Z\ {\rm quaternion\ real} \ear $  & \cr \hline 
{\bf 21a} QC$^*$ & $ \barc q=\pzzm \cr c=\zppmz ,\ \epsilon_c=+ \ear $ & $ \barc  \left( \begin{array}{cc} A & B \cr \mp B^* & \pm A^* \end{array} \right)  \cr A=A^\dagger,\ B=\pm B^T \ear $  & $ \barc  \left( \begin{array}{cc} X & Y  \cr \mp Y^* & X^* \end{array} \right)  
\cr X=-X^\dagger,\ Y=\mp Y^T \ear $  & \cr \hline 
{\bf 21b} QC$^*$ & $ \barc q=\pzzm \cr c=\zppmz ,\ \epsilon_c=- \ear $ & $ \barc  \left( \begin{array}{cc} A & B \cr \pm B^* & \pm A^* \end{array} \right)  \cr A=A^\dagger,\ B=\mp B^T \ear $  & $ \barc  \left( \begin{array}{cc} X & Y  \cr \mp Y^* & X^* \end{array} \right)  \cr X=-X^\dagger,\ Y=\mp Y^T \ear $  & \cr \hline 
\end{tabular}
\end{table}}

{\small
\begin{table}[ht]
\hskip-1.7cm
\caption{Explicit representations of random matrix ensembles with PQC symmetry (continued in Table \ref{tab5'}).
See sections \ref{sec-sym} and \ref{sec-com} for explanations and comments.
\label{tab4}}
\vskip5mm

\hskip-1.5cm
\begin{tabular}{c|c|c|c|c}
\cr \hline 
{\rm Sym.}&
{\rm Implementation} &
{\bf P} &
{\bf K} &
$ \barc {\rm Symmetry}\cr  {\rm class} \ear $ \cr \hline 
{\bf 22} PQC & $ \barc p=\pzzm ,\ q=1\!\!1 \cr c=1\!\!1,\ \epsilon_c=\pm \ear $ & $ \barc  \left( \begin{array}{cc} 0 & A \cr \pm A^T & 0 \end{array} \right)  \cr A=\pm A^* \ear $  & $ \barc  \left( \begin{array}{cc} X & 0 \cr 0 & Y \end{array} \right)  \cr X=X^*=-X^T, \cr Y=Y^*=-Y^T \ear $ & BDI \cr \hline 
{\bf 23} PQC$^*$ & $ \barc p=\pzzm ,\cr q=1\!\!1 \cr c=\isy ,\cr \epsilon_c=\pm \ear $ & $ \barc  \left( \begin{array}{cc} 0 & A \cr \pm \bar{A} & 0 \end{array} \right)  \cr A\ {\rm quaternion\ real} \ear $  & $ \barc  \left( \begin{array}{cc} X & 0 \cr 0 & Y \end{array} \right)  \cr X=-\bar{X},\ Y=-\bar{Y} \cr X, Y\ {\rm quaternion \ real} \ear $ & $ \barc {\rm CII} \cr {\rm (for}\ \epsilon_c=+) \ear $  \cr \hline 
{\bf 24} PQC$^*$ & $ \barc p=\pzzm \cr q=1\!\!1 \cr c=\zppz ,\ \epsilon_c=\pm \ear $ & $ \barc  \left( \begin{array}{cc} 0 & A \cr \pm A^* & 0 \end{array} \right)  \cr A=\pm A^T \ear $  & $ \barc  \left( \begin{array}{cc} X & 0 \cr 0 & X^* \end{array} \right)  \cr X=-X^\dagger \ear $ & $\barc {\rm CI} \cr {\rm DIII} \ear $ \cr \hline 
{\bf 25} PQC$^*$ & $ \barc p=\pzzm \cr q=\zppz  \cr  c=1\!\!1 ,\ \epsilon_c=\pm \ear $ & $ \barc  \left( \begin{array}{cc} 0 & A \cr \pm A^* & 0 \end{array} \right)  \cr A= A^\dagger \ear $  & $ \barc  \left( \begin{array}{cc} X & 0 \cr 0 & X^* \end{array} \right)  \cr X=-X^T \ear $ & \cr \hline 

\end{tabular}
\end{table}}

{\small
\begin{table}[ht]
\caption{Explicit representations of random matrix ensembles with PQC symmetry (continued from Table \ref{tab4} and continued in Table \ref{tab5}).
See sections \ref{sec-sym} and \ref{sec-com} for explanations and comments.
\label{tab5'}}
\vskip5mm

\hskip-2.3cm
\begin{tabular}{c|c|c|c|c}
\cr \hline 
{\rm Sym.}&
{\rm Implementation} &
{\bf P} &
{\bf K} &
$ \!\!\!\barc {\rm Sym.}\cr  {\rm class} \ear $ \cr \hline 

{\bf 26} PQC$^*$ & $ \barc p=\pzzm \cr q=\zppz  \cr  c=\ISY ,\cr \epsilon_c=\pm \ear $ & $ \left( \begin{array}{cc} \!\!\! \!\!\! & \!\!\! \begin{array}{cc} A & B \cr \pm B^\dagger & C \end{array} \!\!\! \cr \!\!\!  \begin{array}{cc}  \pm C^T & \mp B^T \cr  \mp B^* & \pm A^T \end{array} \!\!\!  & \!\!\! \!\!\! \end{array} \right) $  & $ \barc  \left( \begin{array}{cc} \!\!\! \begin{array}{cc} X & Y \cr Z  & -X^T \end{array} \!\!\!  & \!\!\! \!\!\!   \cr \!\!\! \!\!\!  & \!\!\! \begin{array}{cc} -X^\dagger &  -Z^\dagger \cr -Y^\dagger & X^* \end{array} \!\!\!  \end{array} \right)  \cr = \left( \begin{array}{cc} U & 0 \cr 0 & -U^\dagger \end{array} \right); \cr U\equiv \left( \begin{array}{cc} X & Y \cr Z & -X^T \end{array} \right) \ear $ & \cr \hline 
{\bf 27} PQC & $ \barc p=\pzzm  \cr q=\zppz  \cr c=\zppz, \ \epsilon_c=\pm \ear $ & $ \barc  \left( \begin{array}{cc} 0 & A \cr B & 0 \end{array} \right)  \cr A=A^*=\pm A^T,\cr B=B^*=\pm B^T  \ear $ & $ \barc  \left( \begin{array}{cc} X & 0 \cr 0 & -X^T \end{array} \right)  \cr X=X^* \ear $ & \cr \hline 
{\bf 28} PQC & $ \barc p=\pzzm \cr q=\zppz  \cr  c=\XISY ,\cr \epsilon_c=\pm \ear $ 
& $ \barc \left( \begin{array}{cc} \!\!\! \!\!\! & \!\!\! \begin{array}{cc} A & B \cr \mp B^* & \pm A^* \end{array} \!\!\! 
                 \cr \!\!\!  \begin{array}{cc} C & D \cr  \mp D^* & \pm C^* \end{array} \!\!\!  & \!\!\! \!\!\! \end{array} \right)    
               \cr A=A^\dagger,\ B=\mp B^T, \cr C=C^\dagger,\ D=\mp D^T \ear $  
& $ \barc  \left( \begin{array}{cc} \!\!\! \begin{array}{cc} X & Y \cr -Y^*  & X^* \end{array} \!\!\!  & \!\!\! \!\!\!   
                  \cr \!\!\! \!\!\!  & \!\!\! \begin{array}{cc} -X^\dagger &  Y^T \cr -Y^\dagger & -X^T \end{array} \!\!\!  
                  \end{array} \right)  \cr 
                  = \left( \begin{array}{cc} Z & 0 \cr 0 & -Z^\dagger \end{array} \right); \cr 
                  Z \equiv \left( \begin{array}{cc} X & Y \cr -Y^* & X^* \end{array} \right) \ear $ 
                & \cr \hline

\end{tabular}
\end{table}}

{\small
\begin{table}[ht]
\hskip-1.7cm
\caption{Explicit representations of random matrix ensembles with PQC symmetry (continued from Table \ref{tab5'}).
See sections \ref{sec-sym} and \ref{sec-com} for explanations and comments.
\label{tab5}}
\vskip5mm

\hskip-2cm
\begin{tabular}{c|c|c|c|c}
\cr \hline 
{\rm Sym.}&
{\rm Implementation} &
{\bf P} &
{\bf K} &
$ \!\!\!\barc {\rm Sym.}\cr  {\rm class} \ear $ \cr \hline 
{\bf 29} PQC & $ \barc p=\pzzm \cr q=\SZ \cr c=\SX , \cr \epsilon_c=\pm \ear $ & $ \left( \begin{array}{cc} \!\!\! \!\!\! & \!\!\! \begin{array}{cc} A & B \cr \mp B^* & \pm A^* \end{array} \!\!\! \cr \!\!\!  \begin{array}{cc} A^\dagger & \pm B^T \cr  -B^\dagger & \pm A^T \end{array} \!\!\!  & \!\!\! \!\!\! \end{array} \right) $ & $ \barc  \left( \begin{array}{cc} \!\!\! \begin{array}{cc} X & Y \cr -Y^*  & X^* \end{array} \!\!\!  & \!\!\! \!\!\!   \cr \!\!\! \!\!\!  & \!\!\! \begin{array}{cc} U &  V \cr -V^* & U^* \end{array} \!\!\!  \end{array} \right)  \cr X=-X^\dagger , \ Y=-Y^T, \cr U=-U^\dagger ,\ V=-V^T \ear $ & \cr \hline 
{\bf 30} PQC & $ \barc p=\pzzm \cr q=\SZ \cr c=\ISY , \cr \epsilon_c=\pm \ear $ & $ \left( \begin{array}{cc} \!\!\! \!\!\! & \!\!\! \begin{array}{cc} A & B \cr \pm B^* & \pm A^* \end{array} \!\!\! \cr \!\!\!  \begin{array}{cc} A^\dagger & \mp B^T \cr  -B^\dagger & \pm A^T \end{array} \!\!\!  & \!\!\! \!\!\! \end{array} \right) $ & $ \barc  \left( \begin{array}{cc} \!\!\! \begin{array}{cc} X & Y \cr Y^*  & X^* \end{array} \!\!\!  & \!\!\! \!\!\!   \cr \!\!\! \!\!\!  & \!\!\! \begin{array}{cc} U &  V \cr V^* & U^* \end{array} \!\!\!  \end{array} \right)  \cr X=-X^\dagger , \ Y=Y^T, \cr U=-U^\dagger ,\ V=V^T \ear $ & \cr \hline

\end{tabular}
\end{table}}

\subsection{Selected comments on specific ensembles}
\label{sec-spec-com}

In this subsection we comment on specific ensembles.
The comments are not exhaustive; the same type of comments may be valid for other ensembles than
the one commented upon. Comments are labeled with the number of the row in the tables on which we find the ensemble commented upon. We do not give references to literature where ensembles are used,
because it would be impossible to quote all the relevant papers.

{\bf 8} In \cite{BLC} this ensemble was represented by $$k=\zpmz $$ 
Here, the chosen basis corresponds to the one in which the matrix elements are real quaternions. 

{\bf 11} The same ensembles correspond to two different implementations, 
$$p=\pzzm , \ \ \ c=\zppz , \ \ \ \epsilon_c=\pm $$ or equivalently, $$p=\pzzm , \ \ \ c=\zpmz , \ \ \ \epsilon_c=\mp .$$

{\bf 12, 16} The two ensembles 
$\left\{ \left( \begin{array}{cc} 0 & A \cr \pm A^\dagger & 0 \end{array} \right) \right\} $ and the two ensembles 
$\left\{ \left( \begin{array}{cc} 0 & A \cr \pm A^* & 0 \end{array} \right) \right\} $
are simply related by the Weyl unitary trick.

{\bf 14, 16} The two cases on line {\bf 14} are unitarily equivalent to each other, and so are 
the two cases on line {\bf 16}.

{\bf 15}  Another form for this ensemble is given by the unitarily equivalent form defined by
$$ p=\pzzm , \ \ \ k=\ISY $$

This was the form used in \cite{BLC}, but it does not make the quaternion structure evident. 
In this case the subspace ${\bf P}$ is defined by matrices of the form

$$ \left( \begin{array}{cc} \!\!\! \!\!\! & \!\!\! \begin{array}{cc} A & B \cr -B^* & A^* \end{array} \!\!\! \cr \!\!\! \begin{array}{cc} C & D \cr -D^* & C^* \end{array} \!\!\! & \!\!\! \!\!\!  \end{array} \right) $$

and the algebra ${\bf K}$ by

$$ \left( \begin{array}{cc} \!\!\! \begin{array}{cc} X & Y \cr -Y^* & X^* \end{array} \!\!\! &\!\!\! \!\!\!  \cr \!\!\! \!\!\! & \!\!\! \begin{array}{cc} U & V \cr -V^* & U^* \end{array} \!\!\! \end{array} \right) $$

This example shows how different representations can look; however, the above form
does give a hint about the quaternion real basis when one recalls that a real quaternion has the form

$$ q=\left( \begin{array}{cc} z & w \cr -w^* & z^*\end{array} \right)$$ 

(On the other hand, for example ensemble {\bf 28} for $\epsilon_c=+$ looks like it may have such a basis
and it doesn't.)

{\bf 18a, 18b} These two QC--symmetric hermitean ensembles differ by the sign of $\epsilon_c $. However, we have chosen different antisymmetric representations for $c$ in the two cases.
This is possible since $q=1$ so the commutativity constraint for QC is trivial and satisfied for both choices of $c$. 
Thus, the choice $$c=\zppmz $$ for $c$ is also possible in ensemble {\bf 18a};
this choice leads to the alternative form for ${\bf P}$, 
$${\bf P'_{\pm }}=\left\{ \left( \begin{array}{cc} A & B \cr \pm B^* & A^* \end{array} \right): A=A^\dagger, \ B=\pm B^T \right\}$$ 
The ensemble ${\bf P'_-}$ associated with $$c=\zpmz $$ is not expressed in terms of real quaternions, but it is 
related by a unitary transformation to the form given in the table for $$c=\isy $$
The ensemble obtained with $c=\zppz $ on line {\bf 18b} is the one obtained by Altland and Zirnbauer 
for particle--hole symmetric Hamiltonians of Cartan class D in \cite{AZ}. Inspection of its form shows that 
it indeed has the same form as the matrices of this class on line {\bf 17}.

Alternative forms of ensemble {\bf 19} are given in Table \ref{tab-equi20}.

{\small
\begin{table}[ht]
\caption{Unitarily equivalent representations to the ensemble on line {\bf 19}. The signs refer to the sign
alternatives on line {\bf 19} and a prime denotes yet an alternative form (e.g. line {\bf 19$'$} is equivalent to the two ensembles on line {\bf 19}; line {\bf 19$+$}
and {\bf 19$'+$} are equivalent to the upper sign alternative on line {\bf 19}; etc.).
\label{tab-equi20}}
\vskip5mm

\begin{tabular}{c|c|c|c}
\cr \hline 
{\rm Sym.}&
{\rm Implementation} &
{\bf P} &
{\bf K} \cr \hline 

{\bf 19$'$} QC& $\barc q=\pzzm \cr c=\pzzm ,\ \epsilon_c=\pm \ear $ & $ \barc  \left( \begin{array}{cc} A & B \cr \mp B^T & C \end{array} \right)  \cr A=\pm A^*=\pm A^T,\cr B=\pm B^*,\cr C=\pm C^* = \pm C^T  \ear $  & $ \barc  \left( \begin{array}{cc} X & Y  \cr Y^\dagger & Z \end{array} \right)  \cr X=X^*=-X^T,\cr Y=-Y^*,\cr Z=Z^*=-Z^T \ear $  \cr \hline 
{\bf 19$+$} QC & $ \barc q=\zppz \cr c=1\!\!1,\ \epsilon_c=+ \ear $ & $ \barc  \left( \begin{array}{cc} A & B \cr  B^* & A^* \end{array} \right)  \cr A=A^T,\ B=B^\dagger \ear $  & $ \barc  \left( \begin{array}{cc} X & Y  \cr Y^* & X^* \end{array} \right)  \cr X=-X^T,\ Y=-Y^\dagger \ear $   \cr \hline 
{\bf 19$-$} QC & $ \barc q=\zppz \cr c=1\!\!1,\ \epsilon_c=- \ear $ & $ \barc  \left( \begin{array}{cc} A & B \cr -B^* & -A^* \end{array} \right)  \cr A=-A^T,\ B=B^\dagger \ear $  & $ \barc  \left( \begin{array}{cc} X & Y  \cr Y^* & X^* \end{array} \right)  \cr X=-X^T,\ Y=Y^\dagger \ear $   \cr \hline 
{\bf 19$'+$} QC & $ \barc q=\zppz \cr c=\zppz ,\ \epsilon_c=+ \ear $ & $ \barc  \left( \begin{array}{cc} A & B \cr C & A^T \end{array} \right)  \cr A=A^* \cr B=B^*=B^T \cr C=C^*=C^T \ear $  & $ \barc  \left( \begin{array}{cc} X & Y  \cr Z & -X^T \end{array} \right)  \cr X=X^* \cr Y=Y^*=-Y^T \cr Z=Z^*=-Z^T \ear $   \cr \hline 
{\bf 19$'-$} QC & $ \barc q=\zppz \cr c=\zppz ,\ \epsilon_c=- \ear $ & $ \barc  \left( \begin{array}{cc} A & B \cr C & -A^T \end{array} \right)  \cr A=-A^* \cr B=-B^*=-B^T \cr C=-C^*=-C^T \ear $  & $ \barc  \left( \begin{array}{cc} X & Y  \cr Z & -X^T \end{array} \right)  \cr X=X^* \cr Y=Y^*=-Y^T \cr Z=Z^*=-Z^T \ear $   \cr \hline 
\end{tabular}
\end{table}}

Lines {\bf 20a, 20b} describe ensembles ${\bf P}$ of self--dual and anti--selfdual matrices, respectively. 
These ensembles, that have been listed on two separate lines for clarity, are related by the Weyl unitary trick
and correspond to the same symmetric subalgebra.

Unitarily equivalent representations of the ensembles on lines {\bf 21a}, {\bf 21b} 
corresponding to the same ${\bf K}$, are given in Table \ref{tab-equi2425}.

{\small
\begin{table}[ht]
\caption{Unitarily equivalent representations to the ensembles on lines {\bf 21a} and {\bf 21b}.
Ensemble {\bf 21a$+$} is equivalent to the upper sign alternative in 
ensemble {\bf 21a} etc.\label{tab-equi2425}}
\vskip5mm

\begin{tabular}{c|c|c|c}
\cr \hline 
{\rm Sym.}&
{\rm Implementation} &
{\bf P} &
{\bf K} \cr \hline 
{\bf 21a$+$} QC & $ \barc q=\zppz \cr c=\pzzm,\ \epsilon_c=+ \ear $ & $ \barc  \left( \begin{array}{cc} A & B \cr -B^* & A^* \end{array} \right)  \cr A=A^T,\ B=B^\dagger \ear $  & $ \barc  \left( \begin{array}{cc} X & Y  \cr -Y^* & X^* \end{array} \right)  \cr X=-X^T,\ Y=-Y^\dagger \ear $   \cr \hline 
{\bf 21b$+$} QC & $ \barc q=\zppz \cr c=\pzzm ,\ \epsilon_c=- \ear $ & $ \barc  \left( \begin{array}{cc} A & B \cr B^* & -A^* \end{array} \right)  \cr A=-A^T,\ B=B^\dagger \ear $  & $ \barc  \left( \begin{array}{cc} X & Y  \cr -Y^* & X^* \end{array} \right)  \cr X=-X^T,\ Y=Y^\dagger \ear $   \cr \hline 
{\bf 21a$-$} QC & $ \barc q=\zppz \cr c=\zpmz ,\ \epsilon_c=+ \ear $ & $ \barc  \left( \begin{array}{cc} A & B \cr C & A^T \end{array} \right)  \cr A=A^* \cr B=-B^*=-B^T \cr C=-C^*=-C^T \ear $  & $ \barc  \left( \begin{array}{cc} X & Y  \cr Z & -X^T \end{array} \right)  \cr X=X^* \cr Y=-Y^*=Y^T \cr Z=-Z^*=Z^T \ear $   \cr \hline 
{\bf 21b$-$} QC & $ \barc q=\zppz \cr c=\zpmz ,\ \epsilon_c=- \ear $ & $ \barc  \left( \begin{array}{cc} A & B \cr C & -A^T \end{array} \right)  \cr A=-A^* \cr B=B^*=B^T \cr C=C^*=C^T \ear $  & $ \barc  \left( \begin{array}{cc} X & Y  \cr Z & -X^T \end{array} \right)  \cr X=X^* \cr Y=-Y^*=Y^T \cr Z=-Z^*=Z^T \ear $   \cr \hline 

\end{tabular}
\end{table}}

On line {\bf 23} for $\epsilon_c=\pm $, ${\bf P}$ consists of self--dual and anti--selfdual matrices, respectively.

On line {\bf 24} we see that the two Cartan ensembles CI and DIII (i.e., DIII-even in the
notation of \cite{SS}) are dual to each other and correspond to the same symmetric subalgebra\footnote{This may at first sight appear surprising since they correspond to Dyson index $\beta=1$ and $\beta=4$, respectively.}
The two ensembles are {\it not} related by the Weyl trick, and correspond to different
symmetry conditions for the submatrix $A$. 
That these ensembles really correspond to Cartan classes CI and DIII can be verified by comparing the
form of the representative matrices with equations (2.17n$'$) and (2.18n$'$) of Chapter 9 in the book by Gilmore \cite{Gilmore}. Indeed, the two Cartan classes
correspond to the cosets $${\bf SO^*(2n)/U(n,C)}$$ and $${\bf Sp(2n,R)/U(n,C)},$$ respectively, invariant under the 
same symmetric subgroup. 
There are other possible implementations of the matrices $q$ and $c$. These are:

$$ q=1\!\!1, \ \ \ c=\zpmz , \ \ \ \epsilon_c=\mp $$

$$ q=\pzzm , \ \ \ c=\zppz , \ \ \ \epsilon_c=\mp $$

$$ q=\pzzm , \ \ \ c=\zpmz , \ \ \  \epsilon_c=\mp $$

Each of these possibilities corresponds to the same two ensembles given in the table, except that
the last possibility gives rise to ensembles related by the Weyl unitary trick to the original ones. 

On line {\bf 25} an equivalent implementation of $c$ is $$c=\pzzm ,\ \ \ \epsilon_c=\mp .$$

On line {\bf 26} an equivalent implementation of $c$ is $$c=\ISy ,\ \ \ \epsilon_c=\mp .$$

On lines {\bf 29} and {\bf 30}, the ensembles look very similar. However, they are never the same,
because the upper and lower signs correspond to different ensembles. Line {\bf 29} corresponds to a symmetric $c$
and line {\bf 30} to an antisymmetric $c$, even though the commutativity relations between $p$, $q$ and $c$ are
identical for the two entries of the table. They also correspond to different symmetric subalgebras with different symmetries 
of the submatrices $Y$ and $V$. Apparently, these ensembles were forgotten in \cite{BLC}, except for the
equivalences of their commutativity relations that were mentioned on line 2 in the table\footnote{We also believe there is a typing mistake in the last line of the same table which should read
$$(c^T=\pm\epsilon_cc;\ \epsilon_{cp}=\epsilon_{pq}=-\epsilon_{cq}=-)_{\epsilon_c} \simeq
(c^T=\pm\epsilon_cc;\ \epsilon_{cp}=\epsilon_{pq}=\epsilon_{cq}=-)_{\epsilon_c}$$
corresponding to the ensembles on line {\bf 27} and line {\bf 28}.} on p. 5.

\section{Conclusions}

Starting from physically motivated symmetries, we have enumerated a number of hermitean and
non--hermitean random matrix ensembles defining symmetric spaces, giving explicit representations 
of the ensembles and the corresponding symmetry transformations in
terms of Lie algebras. Thus, we have proved that just like ensembles of hermitean hamiltonian, Dirac operator, scattering and transfer matrices of physical systems with given symmetries can be classified 
as symmetric spaces, 
so also a large number of non--hermitean random matrix ensembles define {\it complex} symmetric manifolds. We have tried to include in our tables all the inequivalent ensembles defined by combinations of certain physically motivated symmetries. Our classification is complete in
this sense (but we have not proved there are no other possible random matrix ensembles defining symmetric manifolds). 
 
This constitutes only the first step towards physical applications of the theory of symmetric
spaces to {\it non--hermitean} random matrix ensembles, while such applications can already be found in the
literature for hermitean ensembles, extending all the way to the computation of physical measurable quantities
expressed group theoretically in terms of the underlying symmetric space and its root lattice.
The next step should consist in identifying the non--hermitean ensembles in Tables \ref{tab1} -- \ref{tab5} to 
understand the nature of the corresponding geometrical manifolds. 
We are particularly interested in the operators corresponding to free diffusion on such spaces, as these may
lead to immediate physical applications. 

In conclusion, the symmetric space
picture provides a new approach to non--hermitean random matrix ensembles and puts
them into a unified context with the hermitean ensembles.

\begin{appendix}
\setcounter{equation}{0}
\section{Symmetric spaces from a random matrix viewpoint}
\label{sec-appA}

A random matrix ensemble is defined by a partition function

\beq
Z_\beta \sim \int dH\, P_\beta(H)
\eeq

where $H$ is a random $N\times N$ matrix, $dH$ is an invariant (Haar) measure on the ensemble, and 
$P_\beta(H)$ is a probability measure containing the random matrix potential $V(H)$:

\beq
P_\beta(H)\propto {\rm e}^{-c\beta N{\rm tr} V(H)}
\eeq

where $c$ is a constant. 
The partition function depends on a symmetry index $\beta $, the Dyson index, that takes one of the values 1, 2 or 4
for real, complex or real quaternion--valued matrix elements, respectively. Its value depends on the
presence or absence of time reversal symmetry and space rotation symmetry (or alternatively, time reversal
symmetry and even or odd half--integer spin). $P_\beta (H)$ is often taken 
to have a simple gaussian form, because the correlation functions are to a large extent
independent of the form of the potential $V(H)$ in a certain universal scaling limit, the {\it microscopic limit},
in which the dependence on the average level spacing has been removed. 
Ensembles are invariant under an automorphism of the type

\beq
H\to UHU^{-1}
\eeq

where $U$ is a unitary matrix belonging to a subgroup of
invariance. It is evident that this invariance
property defines a coset structure, which we will show leads to the
identification of each ensemble with a symmetric coset space.

For a given $U$, the random matrix $H$ can be diagonalized:

\beq
\label{eq:diag}
H\to U \Lambda U^{-1}
\eeq

where $\Lambda$ is (block-)diagonal and contains the random matrix
eigenvalues $\{ \lambda_i\} $. Upon performing the similarity transformation 
(\ref{eq:diag}), the random matrix partition function factorizes:

\beq
\label{eq:factorizedZ}
Z_\beta \to \int dU\, \int \prod_i^N d\lambda_i J_\beta (\{ \lambda_i\} )
P_\beta(\{\lambda_i\})
\eeq

This means that the part that depends on the eigenvectors in $U$ can be dropped
by choosing an appropriate normalization of $Z_\beta $, and we are left with an integral only over the eigenvalues.
$J_\beta (\{ \lambda_i\})$ is the Jacobian of the
transformation (\ref{eq:diag}), e.g., for a Wigner--Dyson ensemble with index $\beta$ it has the form
of a product of Vandermonde determinants

\beq
J_\beta (\{ \lambda_i\} ) \sim \prod_{i<j} |\lambda_i-\lambda_j|^\beta 
\eeq

It is the Jacobian that gives rise to the interaction between eigenvalues,
because it produces a logarithmic repulsive pair potential. 
In general however, the Jacobian  does not have this simple form, which is valid only for
the Wigner--Dyson ensembles of random Hamiltonians. It depends
on the type of {\it root lattice} and on the {\it curvature} of the underlying symmetric space. The  general form
of the Jacobian is given by formulas from the theory of symmetric spaces \cite{Helgason}

\beq
\label{eq:J_j}
\begin{array}{l} 
J^0(q)=\prod_{\alpha \in R^+} ({\bf q}\cdot \alpha )^{m_\alpha }\\
\\
J^-(q)=\prod_{\alpha \in R^+} {\rm sinh}^{m_\alpha }({\bf q}\cdot \alpha )\\
\\
J^+(q)=\prod_{\alpha \in R^+} {\rm sin}^{m_\alpha }({\bf q}\cdot \alpha )\end{array}
\eeq 

These expressions deserve some explanation as they are at the heart of the correspondence
between symmetric spaces and random matrices.

In the random matrix partition function, the matrix $H$ belongs to some semisimple Lie group $G$ or, 
in case of a random Hamiltonian or Dirac operator, to the Lie algebra ${\bf G}$ of such a group, while
the matrix $U$ in (\ref{eq:diag}) belongs to a subgroup $K \subset G$ in both cases.
The removal of the unitary subgroup $K$ in (\ref{eq:factorizedZ}) corresponds to 
taking the {\it coset space} $G/K$, or {\it algebra subspace} ${\bf G/K}$ in case $H$ is in the algebra, as the integration
manifold. Both $G/K$ and ${\bf G/K}$ are invariant under conjugation by the subgroup $K$, and the algebra subspace actually has
a larger, non--semisimple invariance group $p\to kpk^{-1}+p'$ which involves also translation, 
i.e. addition of an element from the algebra
${\bf P}\equiv {\bf G/K}$ (we assume that the subgroup $K$ is connected). 

The integration manifold then is a (symmetric) coset space, whose {\it radial} coordinates are associated to the random matrix 
eigenvalues. These radial coordinates are exactly the $({\bf q}\cdot {\bf \alpha})$'s in (\ref{eq:J_j}), where ${\bf q}$ are local coordinates identified with the $\lambda_i$'s and ${\bf \alpha}$ 
are {\it restricted roots}. This can be understood by realizing that each point in the symmetric space is conjugate to some
element from the maximal abelian subalgebra associated to the space. 

A space is said to be {\it symmetric} if there is an involutive automorphism $\sigma \neq 1$ of the Lie
algebra ${\bf G}$ onto itself such that $\sigma^2=1$ and 
$\sigma $ splits the algebra ${\bf G}$ into orthogonal
eigensubspaces ${\bf G}={\bf K}\oplus {\bf P}$ corresponding to the eigenvalues $\pm 1$:

\beq
\sigma (X)=+X\ {\rm for}\ X\in {\bf K},\ \
\sigma (X)=-X\ {\rm for}\ X\in {\bf P}
\eeq

such that 

\beq
\label{eq:commrel}
[{\bf K},{\bf K}]\subset {\bf K},\ \  [{\bf K},{\bf P}]\subset {\bf P},\ \
[{\bf P},{\bf P}]\subset {\bf K}
\eeq

${\bf P}$ spans the tangent space of the coset space $G/K={\rm e}^{\bf P}$.
Suppose ${\bf G}={\bf K}\oplus {\bf P}$ is compact; then ${\bf G^*}={\bf K}\oplus i{\bf P}$ is 
non--compact and defines the dual space $G^*/K={\rm e}^{i{\bf P}}$. By enumerating all the 
involutive automorphisms of  the compact real form of a complex algebra,
defined by the following combination of generators $H_i$ in the 
Cartan subalgebra and raising and lowering operators $E_{\pm \alpha}$

\beq
{\bf K}=\left\{\frac{(E_\alpha - E_{-\alpha})}{\sqrt{2}}\right\},\ \ \
{\bf P}=\left\{iH_i,\frac{i(E_\alpha + E_{-\alpha})}{\sqrt{2}}\right\}
\eeq

(the
corresponding real form has a {\it diagonal} metric, where as a metric on the algebra we use the Killing form), we obtain all the 
(non--compact) symmetric spaces associated to the complex algebra. 
A symmetric space is characterized by its 
restricted root system, which is obtained with respect to the maximal abelian subalgebra in ${\bf P}$.

As was reviewed in \cite{SS}, on a Lie algebra ${\bf G}$ with basis $\{X_i\}$ a metric tensor $g_{ij}$
can be defined by the {\it Killing form} $K(X_i,X_j)$:

\beq
\label{eq:metric}
g_{ij}=K(X_i,X_j)\equiv {\rm tr} ({\rm ad}X_i {\rm ad}X_j)=C^r_{is}C^s_{jr}
\eeq

where ${\rm ad}$ denotes the adjoint representation and $C^i_{jk}$ are the structure constants of the
algebra. 
A theorem due to Cartan guarantees that the Killing form is non--degenerate for a
semisimple algebra. We can also define a curvature tensor and sectional curvature (the same as scalar
curvature in 2d). It is easy to show that due to the commutation relations
(\ref{eq:commrel}), the symmetric manifolds have either positive, negative, or zero curvature. On a symmetric space
with zero curvature, which corresponds to the algebra subspace ${\bf P}$ itself, the Killing form is degenerate. We can use the normal Euclidean metric in this case, $g_{ij}=\delta_{ij}$. 

Let us now discuss equation (\ref{eq:J_j}) in more detail, repeating some of the above statements for clarity.

[i] {\bf The superscript on $J$.}
In eq. (\ref{eq:J_j}), the superscript on $J$ ($0,\ \pm $) denotes the curvature of the symmetric space.
The space whose points are parametrized by the sets of eigenvalues $\{ \lambda_i \}$ of random 
{\it Hamiltonians}, having the structure of {\it generators} in the coset space
${\bf G/K}$ of a Lie algebra, has zero curvature. These zero curvature spaces actually have a
symmetry group larger than $K$, because they are also translationally invariant. Their symmetry
group is non--semisimple, being the semidirect product of $K$ with the abelian subalgebra of translations,
corresponding to addition of an element from the algebra subspace ${\bf P}$ itself.  
A space of random {\it scattering matrices} can (at least in a finite range)
be thought of as the space of matrices defined by $S\approx {\rm e}^{i{\cal H}}$ (where ${\cal H}$ is the Hamiltonian
of the physical system). This corresponds to a space of positive curvature, while the spaces with
negative curvature are {\it transfer matrix} ensembles. Because there is no simple relationship
between  the Hamiltonian and the transfer matrix of a physical system, the ensemble of the transfer 
matrix is not the negative curvature ensemble corresponding to the symmetric
subgroup $K$ of the Hamiltonian and the scattering matrix. It corresponds
to another symmetric space ${G^*}'/K'$.

[ii] {\bf The product over $\alpha \in R^+$.} The product goes over all the positive restricted roots of the
root lattice belonging to the symmetric space. We remind the reader that the roots $\{ \alpha_i \}$ of a
Lie algebra are defined as the weights of the adjoint representation

\beq
[H_i,E_\alpha]=\alpha_i E_\alpha\ \ \ \ \ (i=1,...,r)
\eeq

where $H_i$ is in the Cartan subalgebra and $E_\alpha $ is a raising or lowering operator.
The rank of the algebra $r$ is the dimension of the Cartan subalgebra and the roots 
$\alpha=(\alpha_1,..., \alpha_r)$, being functionals
on the Cartan subalgebra, are $r$--dimensional vectors in the space dual to the same.
The {\it restricted roots} are similar objects, but to define these one takes $H_i$ in the maximal
abelian subalgebra (of dimension $r'\leq r$) in the algebra subspace ${\bf P}$ and $E_\alpha $ in the algebra ${\bf G}$. The 
restricted root system thus obtained can be of an entirely different type from the root lattice
inherited from the complex extension algebra.

[iii] {\bf The quantities ${\bf q}\cdot \alpha$.} The quantities ${\bf q}\cdot \alpha $ are exactly the {\it radial
coordinates} on the symmetric space. In general, if $p$ is a point in the symmetric space $P$,
$p'=kpk^{-1}\in P$ (this follows from the commutation relations for ${\bf K}$ and ${\bf P}$).
If $K$ is compact, every $p\in P$ is conjugate with some element
$h={\rm e}^H$ where $H$ is in the maximal abelian subalgebra of ${\bf P}$:

\beq
p=khk^{-1}\ \ \ \ h={\rm e}^{\sum_i q_iH_i}
\eeq

where $k \in K$ and $h$ is the collective {\it spherical radial} coordinate of the point $p\in P$.
The radial coordinates are obtained as the nonzero diagonal elements of the adjoint representation of the 
diagonal matrix $H={\rm ln}h$ in the maximal abelian subalgebra of ${\bf P}$, because they are the 
eigenvalues in this representation, and they consist of opposite pairs of the form $\pm {\bf q}\cdot \alpha$
where $\alpha $ is a restricted root. The local coordinates ${\bf q}=(q^1,...,q^{r'})$ 
on the symmetric space are identified
with random matrix eigenvalues. The product ${\bf q}\cdot {\bf \alpha}$ takes the form $q_i\pm q_j$,
$q_i$ or $2q_i$ for ordinary, short and long roots, respectively. This is exactly how the random matrix
coordinates appear in the Jacobians of matrix models. 
 
[iv] {\bf The exponents $m_\alpha $.} The exponents in (\ref{eq:J_j}) are the multiplicities of the
restricted roots. We remind the reader that in a general root system there are three types of roots:
long, ordinary and short roots. There is a finite number of possible root systems, because the angle
between two roots is limited to a multiple of $\pi /4$ or $\pi / 6$, which follows from the general theory
of weights and roots.

The correspondence between the roots and the characteristic quantities of a random matrix ensemble also
extends to the Dyson and boundary indices $\beta$, $\alpha \equiv m_s+m_l$ (and thereby also to the 
parameters defining the orthogonal polynomials, see \cite{SS}):

\beq
m_o=\beta,\ \ \ \ \ m_l=\beta-1,\ \ \ \ \ m_s=\beta \nu
\eeq

where $\nu $ is the asymmetry index equal to the number of zero modes of the Dirac operator in chiral matrix models.
This follows from the identification of the Jacobians of the ensembles with (\ref{eq:J_j}). 
 
\setcounter{equation}{0}
\section{Quaternionic ensembles}
\label{sec-appB}

Following reference \cite{Mehta1}, where a general--purpose introduction to the properties of quaternions 
and quaternion determinants can be found, we use the representation 

\beq
1\!\!1,\ \ \ e_1=-i\sigma_2,\ \ \ e_2=-i\sigma_1,\ \ \ e_3=+i\sigma_3
\eeq

where $\sigma_i$ are the Pauli matrices,
for the basic quaternion units. They $e_i$'s satisfy 

\beq
e_1^2=e_2^2=e_3^2=e_1e_2e_3=-1\!\!1
\eeq

A (complex) quaternion is defined as

\beq
\label{eq:cq}
q=q_0+\vec{e}\cdot \vec{q}
\eeq

where $\vec{q}=(q_1,q_2,q_3)$ and $q_0, ..., q_3$ are complex numbers. A complex quaternion has a dual, complex conjugate
and hermitean conjugate, defined respectively by

\beq
\label{eq:qdefs}
\bar{q}=q_0-\vec{e}\cdot \vec{q},\ \ \ q^*=q_0^*+\vec{e}\cdot \vec{q^*},\ \ \ q^\dagger = q_0^*-\vec{e}\cdot \vec{q^*}
\eeq

A real quaternion is defined as  in (\ref{eq:cq}) but with $q_0,\ q_i$ real numbers. 
The complex quaternions do not form a field (that is, the inverse of a complex quaternion may not exist), but the 
real quaternions do \cite{Mehta1}. Therefore we may consider matrices that can be expressed in terms of real quaternion elements. An $N\times N$ matrix with real quaternion elements can be written also as a 
$2N\times 2N$ complex matrix, because each quaternion element is a $2\times 2$ matrix.
The real quaternion structure is evident after a certain choice of basis of this $2N \times 2N$ complex matrix, i.e. 
after a unitary transformation that transforms it into a real quaternion matrix.

The dual of a quaternion matrix $A=[a_{ij}]$, where $a$ is a real quaternion, is defined as 

\beq
\bar{A}=[\bar{a}_{ji}]
\eeq

It follows from definition (\ref{eq:qdefs}) that a real quaternion matrix satisfies $A^\dagger =\bar{A}$.
A symplectic matrix is defined as a matrix for which $A\bar{A}=1\!\!1$. 

We note in Tables \ref{tab1} through \ref{tab5} that for the ensembles that can be expressed in terms
of real quaternions, the antisymmetric form 

\beq
\left(\begin{array}{ccccc} 0 &1&&& \cr -1 & 0 &&& \cr && \ddots && \cr &&& 0&1\cr &&& -1&0 \end{array}\right)
=\left(\begin{array}{cccc} -e_1 &&& \cr & -e_1 && \cr && \ddots & \cr &&& -e_1\end{array}\right)
\eeq

of the $k$-- or $c$--matrix defining the symmetry condition makes this structure evident. This can easily be achieved by rewriting the $2\times 2$ 
complex submatrices of the ensemble as complex quaternions and observing that for any quaternion $q$

\beq
-e_1q^Te_1=\bar{q} , \ \ \ -e_1 q^{(*)}e_1=q^* \ \ \ \ \ \ \ (q^{(*)}\equiv q_0^*+\vec{e^*}\cdot \vec{q^*})
\eeq

in the C-- or K--symmetry relations. 
\end{appendix}

 \end{document}